\documentclass[useAMS,usenatbib]{mn2e}
\usepackage{epsfig}
\usepackage{float}
\usepackage{placeins}
\usepackage{graphicx}
\usepackage{epsfig}
\usepackage{bm}
\usepackage{amsfonts}
\usepackage{amssymb}
\usepackage{times}
\usepackage{natbib}
\usepackage{color}

\title[Optical variability of three TeV Blazars]{Multi-band optical variability of three TeV Blazars on Diverse Timescales}

\author[Gupta et al.]
{Alok C.\ Gupta$^{1}$\thanks{E-mail: acgupta30@gmail.com},
A. Agarwal$^{1}$,
J. Bhagwan$^{1,2}$,
A. Strigachev$^{3}$,
R. Bachev$^{3}$,
\newauthor E. Semkov$^{3}$,
H. Gaur$^{4}$,
G. Damljanovic$^{5}$,
O. Vince$^{5}$,
Paul J. Wiita$^{6}$
\\ 
$^{1}$Aryabhatta Research Institute of Observational Sciences (ARIES),
Manora Peak, Nainital -- 263002, India\\
$^{2}$School of Studies in Physics \& Astrophysics, Pt Ravishankar Shukla University, Amanaka G.E. Road, Raipur 492010, India \\
$^{3}$Institute of Astronomy and National Astronomical Observatory,
Bulgarian Academy of Sciences, 72 Tsarigradsko Shosse Blvd., 1784 Sofia, Bulgaria \\
$^{4}$Key Laboratory for Research in Galaxies and Cosmology, Shanghai Astronomical Observatory,
Chinese Academy of Sciences, 80 Nandan Road, Shanghai 200030, China \\
$^{5}$Astronomical Observatory, Volgina 7, 11060 Belgrade, Serbia \\
$^{6}$Department of Physics, The College of New Jersey, 2000 Pennington Rd., Ewing, NJ 08628-0718, USA\\
}

\begin{document}

\date{Accepted ....... Received  ......; in original form ......}

\pagerange{\pageref{firstpage}--\pageref{lastpage}} \pubyear{2014}

\maketitle

\label{firstpage}

\begin{abstract}

We present our optical photometric observations of three TeV blazars, PKS 1510-089, PG 1553+113 and Mrk 501 taken using
two telescopes in India,
one in Bulgaria, one in Greece and one in Serbia during 2012 - 2014.  These observations covered a total of 95 nights with a total of  202 B filter frames,
247 images in V band,
817  in R band
while 229 images were taken in the I filter.
This work is focused on multi-band flux and colour variability studies of these blazars on
diverse timescales which are useful in understanding the emission mechanisms.  We studied the variability characteristics of above three blazars
and found all to be active over our entire observational campaigns.
We also searched for any correlation between
the brightness of the sources and their colour indices.
During the times of variability, no significant evidence for the sources to display spectral changes correlated 
with magnitude was found on timescales of a few months.
We briefly discuss the possible physical mechanisms most likely responsible for the observed flux variability.

 \end{abstract}
 
 \begin{keywords}
 galaxies: active – BL Lacertae objects: general – BL Lacerate objects: individual:
PKS 1510-089 – BL Lacerate objects: individual: PG 1553+113 – BL Lacerate objects: individual: Mrk 501
\end{keywords}

\section{Introduction}

Flat spectrum radio quasars (FSRQs) along with BL Lacertae objects (BL Lacs) constitute the most extreme subclass of
radio-loud Active Galactic Nuclei (RLAGNs) called  blazars, with observed luminosity outshining their host galaxies.
The dominant radiation sources in these objects are believed to originate from relativistic jets that are very strongly
Doppler boosted by being viewed at angles
$\lesssim$ 10$^{\circ}$ with the observer's line of sight (LOS)
(e.g.\ Urry \& Padovani 1995).  This  gives rise to flux variability in all electromagnetic (EM) bands, from radio to very high energy (VHE)
gamma-rays, strong optical ($>$3\%) and radio polarization as well as superluminal motion and anisotropic radiation.

The variability of blazars may either arise from accretion disc (AD) instabilities or changes intrinsic to the relativistic jet,
although  interstellar scintillation can also
cause observed radio variations (e.g.\ Heeschen et al.\ 1987). These fluctuations have been observed at all accessible timescales ranging from a few  minutes through
days and months to even decades. Blazar variability is often broadly divided into three temporal classes. Changes of few hundred to few tenths of magnitude
observed in the flux of a source from minutes to less than a day are categorized as intra-day variability (IDV) or or micro-variability or intra-night variability
(e.g.\ Kinman 1975; Wagner \& Witzel 1995, Clements, Jenks, \& Torres 2003; Rector \& Perlman 2003; Xie et al.\ 2004; Gupta et al.\ 2008a), while variations taking
place from several days to few months are usually known as
short time variability (STV) and those taking from several months to many years are usually called long term variability (LTV; Gupta et al.\ 2004);
in the latter two classes blazar variability often can  exceed even 5 magnitudes.

\begin{table*}
\caption{ Details of telescopes and instruments}
\textwidth=6.0in
\textheight=9.0in
\vspace*{0.2in}
\noindent
\begin{tabular}{p{2.2cm}p{2.2cm}p{2.3cm}p{2.2cm}p{2.3cm}p{2.3cm}p{2.5cm}} \hline
Site:              &A                               &B                             &C                                            & D                                    &E    \\\hline
Scale:           &       0.535\arcsec/pixel       &0.37\arcsec/pixel               &0.2825\arcsec/pixel                             & 0.330\arcsec/pixel$^{\rm a}$        & 0.465\arcsec/pixel                   \\
Field:           & $18\arcmin\times18\arcmin$     & $13\arcmin\times13\arcmin$     &$9.6\arcmin\times9.6\arcmin$      & $16.8\arcmin\times16.8\arcmin$                 &     15.8'x15.8' \\
Gain:            & 1.4 $e^-$/ADU                  &10 $e^-$/ADU                    &2.687 $e^-$/ADU                                    & 1.0 $e^-$/ADU                  &    1.25   $e^-$/ADU           \\
Read Out Noise:  & 4.1 $e^-$ rms                  &5.3 $e^-$ rms                   &8.14 $e^-$ rms                                   & 8.5 $e^-$ rms                    &     3.75  $e^-$ rms                \\
Typical seeing : &   1.2\arcsec to 2.0\arcsec     & 1\arcsec to 2.8\arcsec         & 1\arcsec to 2\arcsec                        & 1.5\arcsec to 3.5\arcsec             &    1\arcsec to 2\arcsec          \\\hline
\end{tabular} \\
A  : 1.30 meter Ritchey-Chretien Cassegrain optical telescope, ARIES, Nainital, India \\
B  : 1.04 meter Sampuranand Telescope, ARIES, Nainital, India  \\
C  : 1.3-m Ritchey-Chretien telescope at Skinakas Observatory, University of Crete, Greece \\
D  : 60-cm Cassegrain telescope at Astronomical Observatory Belogradchik, Bulgaria \\
E  : 60-cm Cassegrain telescope, Astronomical Station Vidojevica - ASV  \\
$^{\rm a}$ With a binning factor of $1\times1$
\end{table*}

Blazar spectral energy distributions (SEDs) consist of two different components (e.g.\ Mukherjee et al.\ 1997; 
Weekes 2003).
At lower frequencies (radio through UV or X-rays) their energy spectrum is widely accepted to be dominated by synchrotron emission from the
relativistic electrons spiraling in the magnetic field of the jets, while most X-rays and all gamma-rays are believed to be explained by inverse Compton (IC)
scattering of low energy photons originating from the AD or  broad line region; however, in most cases the seed photons can be synchrotron photons,
thus producing synchrotron self-Compton emission.
Based on the location of the first peak of the SED
blazars are sub-classified into low energy peaked blazars (LBLs) and high energy peaked blazars (HBLs) with
first hump peaking in the NIR/optical in case of LBLs 
and the second one at GeV energies. In HBLs the first 
peak lies in UV/X-ray band and the second one at TeV $\gamma-$rays energies (e.g.\ Padovani \& Giommi 1995; Abdo et al.\ 2010). 
According to Fossati et al.\ (1997), HBLs are lower luminosity objects with a higher space density and 
stronger magnetic fields, while LBLs are higher luminosity objects, with a lower space density possessing weaker magnetic fields.
To be more quantitative,  the ratio of X-ray flux in the 0.3-3.5 kev band to the radio flux density at 5 GHz, has
been employed, with low-synchrotron-peaked (LSPs) objects
defined as those with this ratio below 10$^{-11.5}$ while for in high-synchrotron-peaked blazars (HSPs) it is greater 
(Padovani \& Giommi 1996). Another class of blazars with SED peaks at intermediate frequencies are sometimes
distinguished and are known as
intermediate synchrotron peaked blazars (ISPs; Sambruna, Maraschi \& Urry 1996).

Miller et al.\ (1989) found the first clear evidence of optical IDV. Carini (1990) found IDV in more than 80\% of blazars when the duration of observations
exceeded 8 hrs. Later, Gupta \& Joshi (2005) studied a larger sample of AGNs and detected significant IDV in $\sim$ 10\% of
radio-quiet Active Galactic Nuclei (RQAGNs), 35--40\% of RLAGNs (excluding blazars) when observed for $\sim$ 6 hours while 
IDV was seen in 80--85\% of blazars when observed duration exceeded 6 hours.
To understand the complex flux variation phenomenon, both quasi-continuous measurements over single nights
and longer term photometric monitoring are required.
Multi-band variability studies help to reveal the true nature of blazars and provide valuable constraints in the emission models.
From long term studies in the optical part of the EM spectra, HBLs are found to be less variable and polarized than LBLs (Jannuzi et al.\ 1994).
They also found the amplitudes of variability to be much smaller in case of HBLs than those for LBLs.
Over the past decade, many new high energy, TeV blazars have been discovered, with a recent number reaching 54 (Holder 2012; 2014),
among which 80\% (44 out of 54) are found to be HBLs. These TeV HBLs  are characterized by strong variability
on diverse timescales ranging from a few months down to even a few minutes (e.g.\ Begelman et al.\ 2008; Nalewajko et al.\  2011;
Gaur, Gupta, \& Wiita 2012; Barkov et al.\ 2012).  Variability studies help us to understand the particle acceleration mechanisms taking place in
the relativistic jets and can  shed light on the central regions of blazars, accretion processes and the structure of the jets.
Rapid variability of TeV blazars at diverse timescales can be more easily observed by virtue of the high bulk Lorentz factors
(often $\geq$ 25) of the knots seen in their relativistic jets.

The key motivation of this paper is to study flux and colour variability characteristics along with spectral changes
of three TeV blazars at diverse timescales
to increase our current understanding of blazar variability through photometric studies.
Here we report the photometric observations of PKS 1510-089, PG 1553+113 and Mrk 501 on diverse timescales
using five telescopes, starting in June 2012 and running through September 2014.
This paper is organized as follows:
in section 2, we describe the observations and data reduction; section 3 gives the details about the analysis techniques used; 
section 4 provides our results of IDV and STV/LTV along with colour variability of our sample of blazars; we present a discussion and our
conclusions in section 5.

\begin{table}
\caption{Observation log of optical photometric observations of PKS 1510-089}
\textwidth=7.0in
\textheight=10.0in
\vspace*{0.2in}
\noindent
\begin{tabular}{lcc} \hline
 Date of Observation & Telescope  & Data Points \\
(yyyy mm dd)         &            &(B, V, R, I) \\\hline

2014 04 30          &D        &0,2,2,2 \\ 
2014 05 20          &D        &0,2,2,2 \\              
2014 05 21          &D        &0,2,2,2 \\
2014 05 22          &D        &0,2,2,2 \\
2014 05 23          &D        &0,2,2,2 \\
2014 06 15          &C        &3,3,3,3  \\               
2014 07 01          &D        &0,2,2,2 \\              
2014 07 02          &D        &0,2,2,2 \\              
2014 07 03          &D        &0,2,2,2 \\ 
2014 07 04          &D        &0,2,2,2 \\
2014 07 05          &D        &0,2,2,2 \\
2014 07 06          &C        &3,3,3,3  \\  
2014 07 21          &C        &3,3,3,3  \\               
2014 07 22          &C        &3,3,3,3  \\               
2014 07 25          &C        &3,3,3,3  \\               
2014 07 29          &C        &3,3,3,3  \\              
2014 08 03          &D       &0,2,2,2 \\ 
2014 08 18          &D        &0,2,2,2 \\ \hline

\end{tabular}
\end{table}

\section{\bf Observations and Data Reductions}

The optical photometric observations of our blazar sample were carried out in the B, V, R, and I pass-bands,
with  two telescopes in India, one in Greece, one in Bulgaria and one in Serbia, all equipped with CCD detectors.
The details of these five telescopes, detectors and other parameters used are given in Table 1.
Over 1495 image frames covering 95 nights between June 2012 and Sept 2014 were taken for three blazars, PKS 1510-089, PG 1553+113 and Mrk 501,
to study their flux and spectral characteristics on
diverse timescales.  Observation logs, including date of observation, number of images acquired
in each filter and the telescope used are presented for the three blazars individually in Tables 2, 3 and 4.

\begin{table}
\caption{Observation log of optical photometric observations of PG 1553+113.}
\textwidth=7.0in
\textheight=10.0in
\vspace*{0.2in}
\noindent
\begin{tabular}{lcc} \hline
 Date of Observation & Telescope  & Data Points \\
(yyyy mm dd)         &            &(B, V, R, I) \\\hline              

 2013 07 03          &E        &1,1,1,1 \\ 
 2013 07 08          &E        &1,1,1,1 \\
 2013 07 26          &C        &3,3,3,3 \\
 2013 08 28          &C        &3,3,3,3 \\
 2014 04 09          &B        &0,1,1,1 \\
 2014 04 10          &B        &1,1,1,1 \\ 
 2014 04 11          &B        &0,1,5,1 \\ 
 2014 04 22          &A        &2,2,33,1 \\ 
 2014 04 23          &A        &1,1,46,1 \\ 
 2014 05 09          &B        &1,1,65,1 \\ 
 2014 05 11          &B        &0,16,149,2 \\ 
 2014 05 17          &B        &1,1,67,1 \\ 
 2014 05 23          &A        &1,1,172,1 \\ 
 2014 05 23          &D        &2,2,2,2 \\
 2014 05 26          &B        &1,1,55,1 \\
 2014 05 27          &E        &3,3,3,3 \\
 2014 06 15          &C        &3,3,3,3 \\
 2014 07 01          &D        &2,2,2,2 \\
 2014 07 02          &D        &2,2,2,2 \\
 2014 07 03          &D        &2,2,2,2 \\
 2014 07 04          &D        &2,2,2,2 \\
 2014 07 05          &D        &2,2,2,2 \\
 2014 07 06          &C        &3,3,3,3 \\
 2014 07 21          &C        &3,3,3,3 \\
 2014 07 22          &C        &3,3,3,3 \\
 2014 07 25          &C        &3,3,3,3 \\
 2014 07 29          &C        &3,3,3,3 \\
 2014 08 18          &D        &2,2,2,2 \\
 2014 08 26          &D        &2,2,2,2 \\

             \hline

\end{tabular}
\end{table}

\begin{table}
\caption{Observation log of optical photometric observations of Mrk 501.}
\textwidth=7.0in
\textheight=10.0in
\vspace*{0.2in}
\noindent
\begin{tabular}{lcc} \hline
 Date of Observation & Telescope  & Data Points \\
(yyyy mm dd)         &            &(B, V, R, I) \\\hline                            

2012 06 02          &C        &3,3,3,3 \\
2012 06 29          &C        &3,3,3,3 \\
2012 06 30          &C        &3,3,3,3 \\ 
2012 07 01          &C        &3,3,3,3 \\ 
2012 07 02          &C        &3,3,3,3 \\ 
2012 07 03          &C        &3,3,3,3 \\ 
2012 07 04          &C        &3,3,3,3 \\ 
2012 07 05          &C        &3,3,3,3 \\ 
2013 07 08          &E        &1,1,1,1 \\ 
2013 07 13          &E        &1,1,1,1 \\ 
2013 07 14          &E        &1,1,1,1 \\  
2013 07 26          &C         &3,3,3,3 \\               
2013 08 28          &C        &3,3,3,3 \\               
2014 03 31          &E        &3,3,3,3  \\               
2014 05 22          &D                 &0,2,2,2 \\
2014 05 23          &D                 &2,2,2,2 \\
2014 05 27          &E        &2,3,2,3  \\               
2014 05 28          &E        &3,3,3,2  \\               
2014 05 28          &E        &5,4,5,4  \\
2014 06 15          &C          &3,3,3,3 \\
2014 06 29          &E        &3,3,3,3  \\               
2014 06 30          &E        &3,3,3,3  \\               
2014 07 01          &E        &3,3,3,2  \\              
2014 07 01          &D        &2,2,2,2 \\
2014 07 02          &D        &2,2,2,2 \\
2014 07 02          &E        &3,2,3,2  \\                             
2014 07 03          &D        &2,2,2,2 \\
2014 07 03          &E        &1,2,3,3  \\               
2014 07 04          &D        &2,2,2,2 \\
2014 07 05          &D        &2,2,2,2 \\ 
2014 07 05          &E        &2,3,3,3  \\               
2014 07 06          &E        &3,3,3,1  \\
2014 07 06          &C        &3,3,3,3  \\              
2014 07 21          &C        &5,5,5,5  \\              
2014 07 22          &C        &5,5,5,5  \\              
2014 07 25          &C        &5,5,5,5  \\              
2014 07 28          &C        &5,5,5,5  \\              
2014 07 29          &C        &5,5,5,5  \\              
2014 08 02          &D        &2,2,2,2 \\ 
2014 08 03          &D        &2,2,2,2 \\               
2014 08 04          &D        &2,2,2,2 \\               
2014 08 18          &D        &2,2,2,2 \\
2014 08 19          &D        &2,2,2,2 \\
2014 08 25          &D        &2,2,2,2 \\
2014 08 26          &D        &2,2,2,2 \\
2014 08 31          &D        &6,6,6,6 \\              
2014 09 18          &D        &2,2,2,2 \\               
2014 09 19          &D        &2,2,2,2 \\                
            \hline

\end{tabular}
\end{table}

\subsection{\bf Telescopes and Data Reduction}
The optical photometric observations of these blazars were carried out using
five telescopes around the world among which
two telescopes are in India operated by
Aryabhatta Research Institute of observational sciencES (ARIES), Nainital. One is
the 1.04 m Sampuranand telescope having Ritchey-Chretien (RC) optics with a f/13
beam equipped with Johnson UBV and Cousins RI filters.  The other is
the 1.3-m Devasthal fast optical telescope (DFOT), which is a fast beam 
(f/4) telescope with a modified RC system equipped with broad band Johnson-Cousins B, V, R, I filters. DFOT 
provides a pointing accuracy better than 10 arcsec RMS (Sagar et al.\ 2011). Further details of both telescopes 
are given in Table 1 (telescopes A and B).
We also employed the 1.3m RC telescope of Skinikas 
Observatory\footnote {Skinakas Observatory is a collaborative project of the University of
Crete, the Foundation for Research and Technology -- Hellas, and the Max-Planck-Institut f\"ur 
Extraterrestrische Physik.}, of the University of Crete (Greece).
Technical parameters and chip specifications for the cameras used are given in Table 1 (Telescope C).
All frames were exposed through a set of standard Johnson-Cousins filters.
In addition to above telescopes, we carried out photometric observations of the blazars using the 60 cm 
Cassegrain telescope of Belogradchik AO, which was equipped with standard UBVRI filter sets.
Instrumental details are summarized in Table 1 (Telescope D).
Observations were also taken with 60cm Cassegrain telescope, which is located on Vidojevica mountain in South Serbia,
through Johnson-Cousins BVRI standard filter set (Telescope E).

IRAF\footnote{IRAF is distributed by the National Optical Astronomy Observatories, which are operated
by the Association of Universities for Research in Astronomy, Inc., under cooperative agreement with the
National Science Foundation.} packages were used for the pre-processing of the raw data following the steps described below.
Bias frames were taken for each night at regular intervals. Taking the median of all bias frames
acquired during a particular night, a master bias was generated which was subtracted from all twilight flat frames and also from the
image frames taken during that night.  The next step was to generate a master flat for each filter by median combining all the flat frames in a particular
passband. Then  each source frame is divided by normalized master flat to remove pixel to pixel inhomogenities.
The final step of image pre-processing is to remove cosmic rays from all source image frames.
Every science exposure 
was bias subtracted, dark subtracted and twilight flat fielded. 
Further processing was then done using the Dominion Astronomical
Observatory Photometry (DAOPHOT II) software (Stetson 1987; Stetson 1992) to perform concentric circular aperture photometry and some
customized scripts written in MATLAB were also used.
Aperture photometry for the data obtained from telescope E, was done with MaxIm DL packages.
For every night aperture photometry was carried out with four different aperture radii, i.e., $\sim 1 \times$~FWHM, $2 \times$~FWHM, 
$3 \times$~FWHM and $4 \times$~FWHM. Aperture radii of $2 \times$~FWHM were finally adopted for our final results as they provided the best SNR.
For these blazars we observed three or more comparison stars from the same field as of the
source\footnote{http://www.lsw.uni-heidelberg.de/projects/extragalactic/charts/}. We then selected two steady
comparison stars based on their proximity in magnitude and colour to the blazar.

\begin{table*}
\caption{ Results of IDV observations of PG 1553+113.} 
\textwidth=7.0in
\textheight=10.0in
\vspace*{0.2in}
\noindent

\begin{tabular}{cccccccc} \hline \nonumber

 Date       & Band   &N          & F-test  &$\chi^{2}$test  &   Variable    &A\% \\
               &                     &                               &$F_{1},F_{2},F,F_{c}(0.99),F_{c}(0.999)$ &$\chi^{2}_{1},
\chi^{2}_{2},\chi^{2}_{av}, \chi^{2}_{0.99}, \chi^{2}_{0.999}$  & & \\\hline 

 22.04.2014   & R     & 30  & 2.08, 1.49, 1.79, 1.20, 1.27 & 38.55, 26.45, 32.5, 746.39, 776.91  & PV  & 3.0 \\
 23.04.2014   & R     & 45  & 2.17, 2.63, 2.40, 1.28, 1.39 & 68.22, 77.19, 72.70, 413.39, 436.37 & PV & 1.0 \\
 09.05.2014   & R     & 56  & 3.27, 3.23, 3.25, 1.89, 2.34 & 125.16, 103.19, 114.17, 82.29, 93.17 & V  &  4.2 \\
 11.05.2014   & R     & 123 & 1.41, 1.56, 1.49, 1.53, 1.76 & 98.46, 110.04, 104.25, 161.25, 176.01 & NV  & -- \\
 17.05.2014   & R     & 67  & 1.32, 1.30, 1.31, 1.27, 1.38  & 48.44, 45.80, 47.12, 37.29, 460.90  & PV  & 4.3 \\             
  22.05.2014  & R     & 170 & 1.08, 1.00, 1.02, 1.73, 2.08 & 115.87, 101.75, 108.81, 104.01, 116.09 & PV  & 4.0 \\
 26.05.2014   & R     & 50  & 2.06, 2.18, 2.12, 1.96, 2.46  & 75.99, 70.10, 73.04, 74.92, 85.35  & PV  & 4.4 \\ 
 
\hline
\end{tabular} \\
\noindent
Var : Variable, PV : probable variable, NV : Non-Variable     \\
\end{table*}

\begin{figure}
\epsfig{figure= 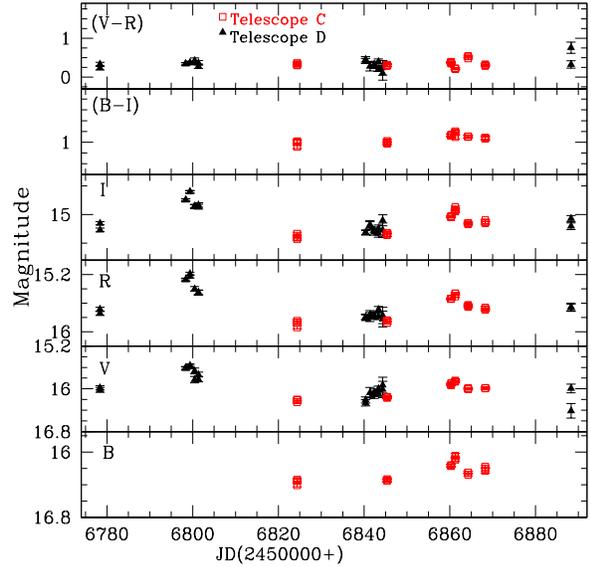,height=3.2in,width=3.1in,angle=0}
\caption{Short-term through long-term variability LCs and colour indices of PKS 1510$-$089 in the B, V, R and I bands 
and (B-I) and (V-R) colours. Different symbols denote data from different observatories: open red squares, telescope C;
 filled black triangles, telescope D. }
\end{figure}

\begin{figure}
\epsfig{figure= 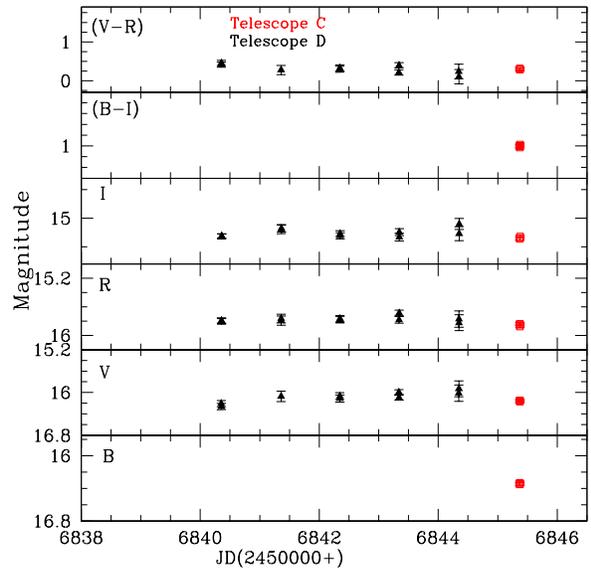,height=3.2in,width=3.1in,angle=0}
\caption{ As in Figure 1, for the limited period around MJD 6840, providing  a better view of variability in that period.}
\end{figure}

\section{\bf Variability detection criterion}
To investigate the IDV properties of blazars we used two statistics, namely the F-test
and $\chi^{2}$ test which helped us to state the statistically significance of the extracted results.

\subsubsection{\bf F-Test}
The F-test is considered to be a proper statistic to determine
any changes of variability. $F$ values compare two sample variances and are calculated as (e.g.\  Agarwal et al.\ 2015):

\begin{equation}
 \label{eq.ftest}
 F_1=\frac{Var(BL-Star A)}{Var(Star A-Star B)}, \nonumber \\ 
 F_2=\frac{Var(BL-Star B)}{ Var(Star A-Star B)}.
\end{equation}

Here (BL-Star A), (BL-Star B), and (Star A-Star B) are the differential instrumental magnitudes of blazar and comparison star A, blazar 
and comparison star B, and star A and star B, respectively, while Var(BL-Star A), Var(BL-Star B), and Var(Star A-Star B) are 
the variances of those differential instrumental magnitudes.

We take the average of $F_1$ and $F_2$ to find a mean observational $F$ value.
The $F$ value is then compared with $F^{(\alpha)}_{\nu_{bl},\nu_*}$, a critical value, where $\nu_{bl}$ and $\nu_*$ 
respectively denote the number of degrees of freedom for the blazar and star, while $\alpha$ is the significance level set as
0.1 and 1 percent (i.e.\ $3 \sigma$ and $2.6 \sigma$) for our analysis. If the mean F value is 
larger than the critical value, the null hypothesis (i.e., that of no variability) is discarded.

\begin{figure*}
\epsfig{figure=  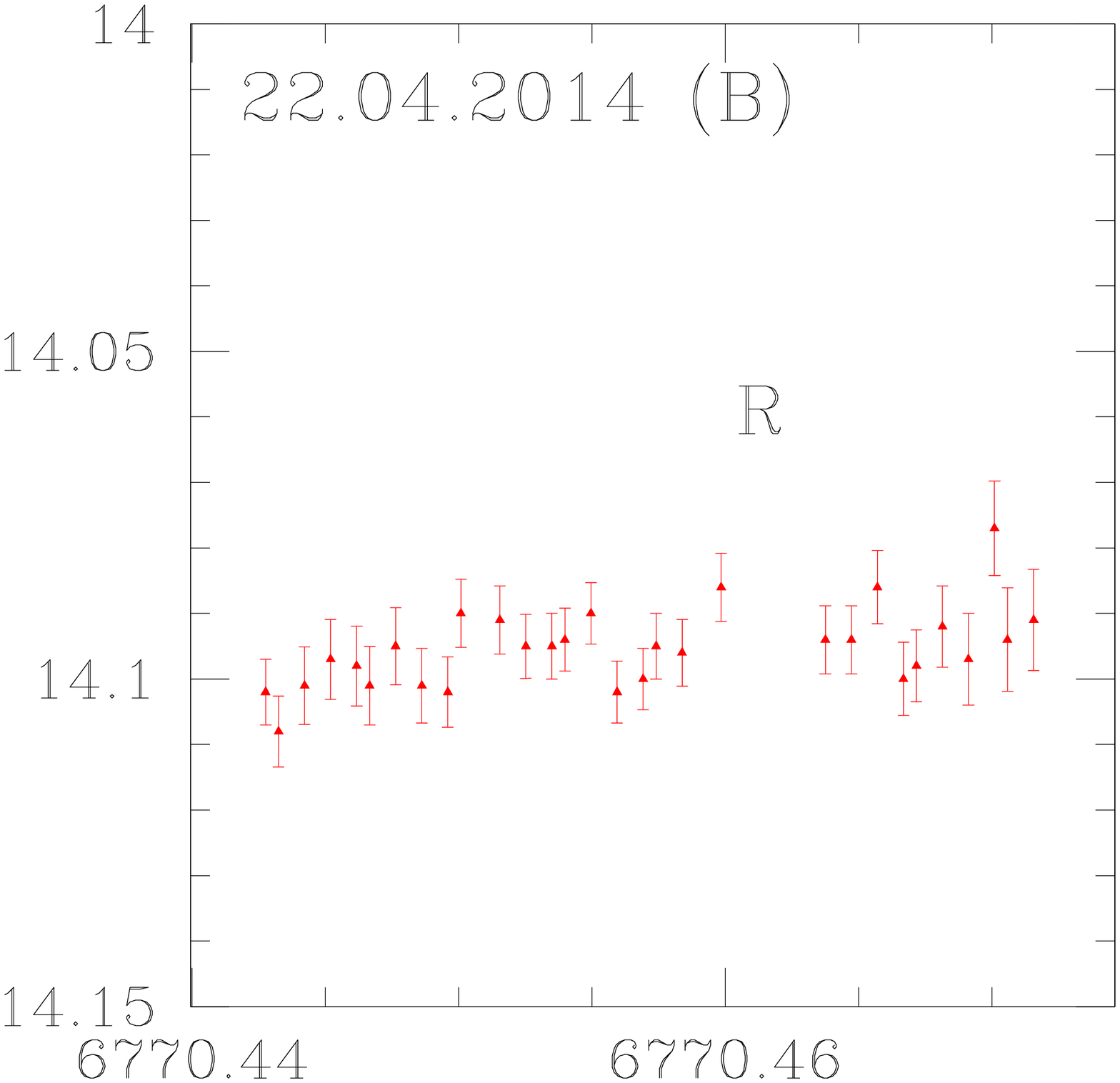,height=1.567in,width=1.59in,angle=0}
 \epsfig{figure=  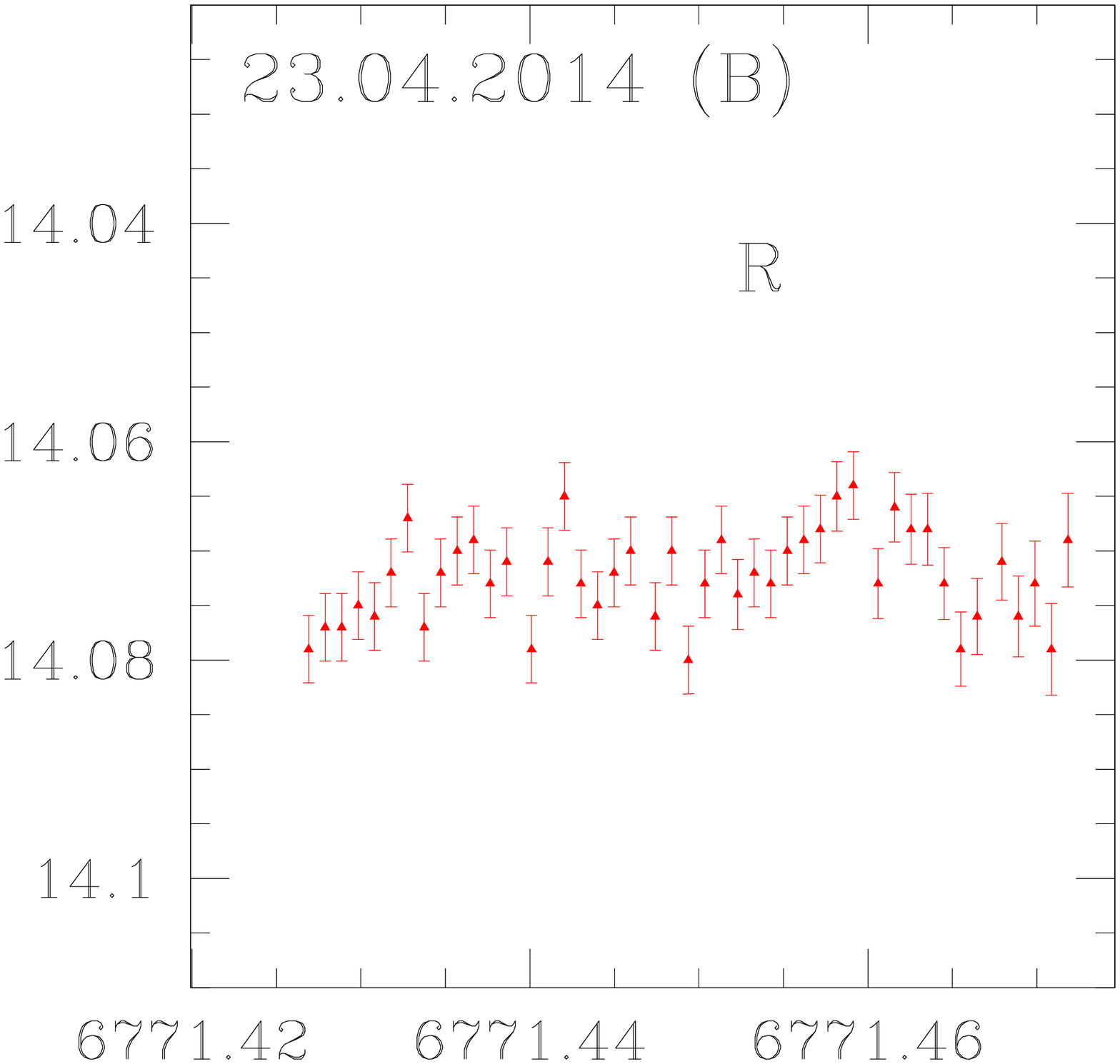,height=1.567in,width=1.59in,angle=0}
\epsfig{figure=  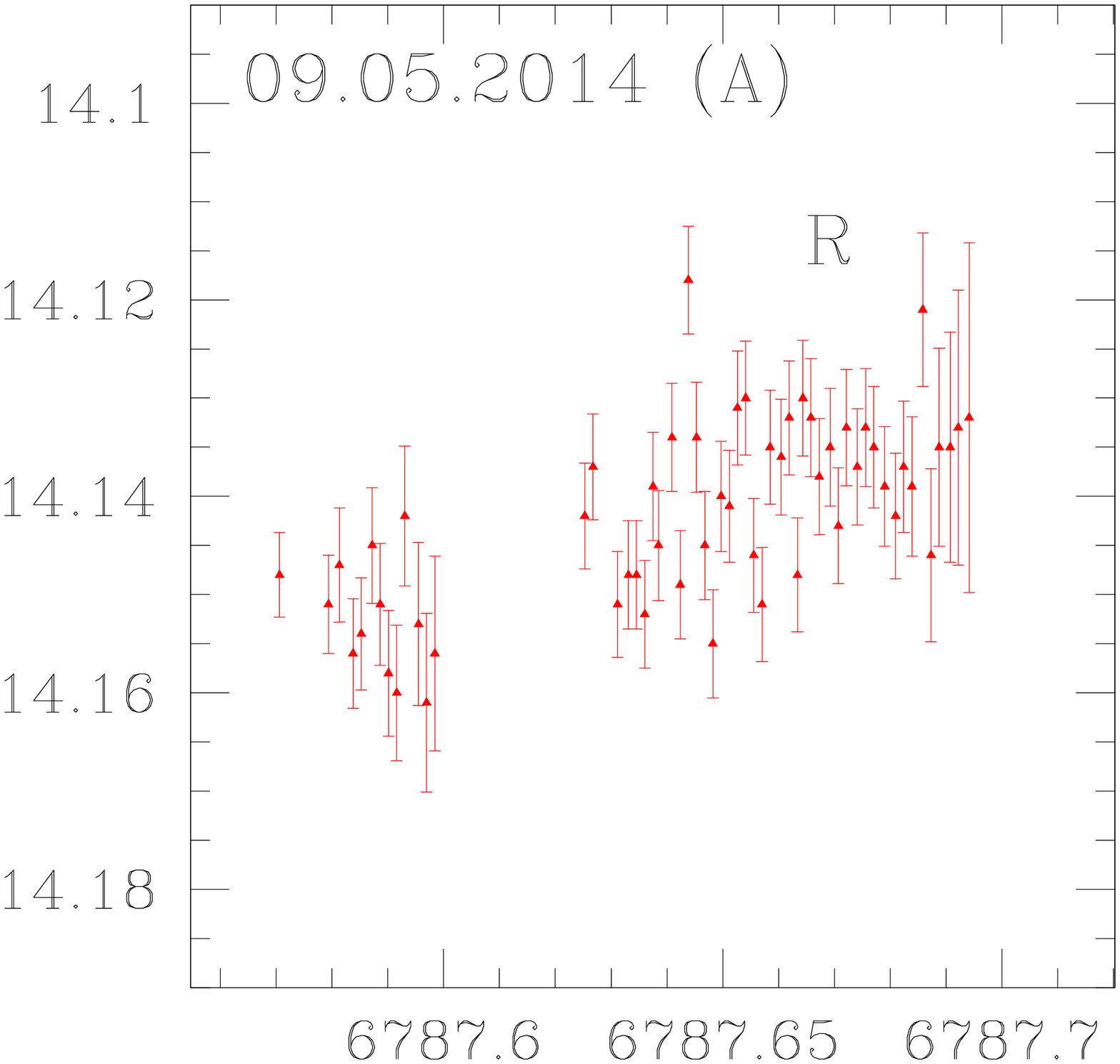,height=1.567in,width=1.59in,angle=0}
\epsfig{figure=  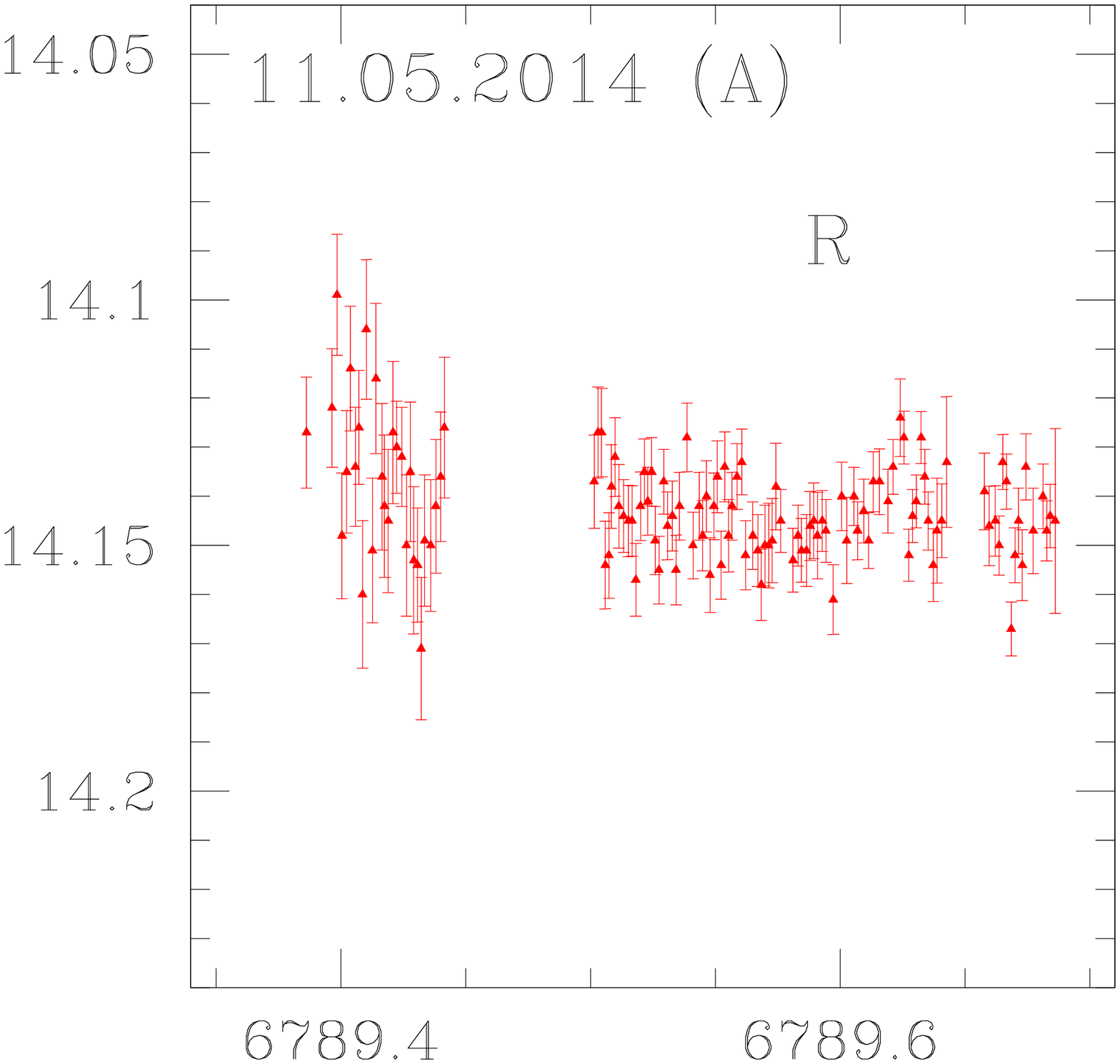,height=1.567in,width=1.59in,angle=0}
\epsfig{figure=  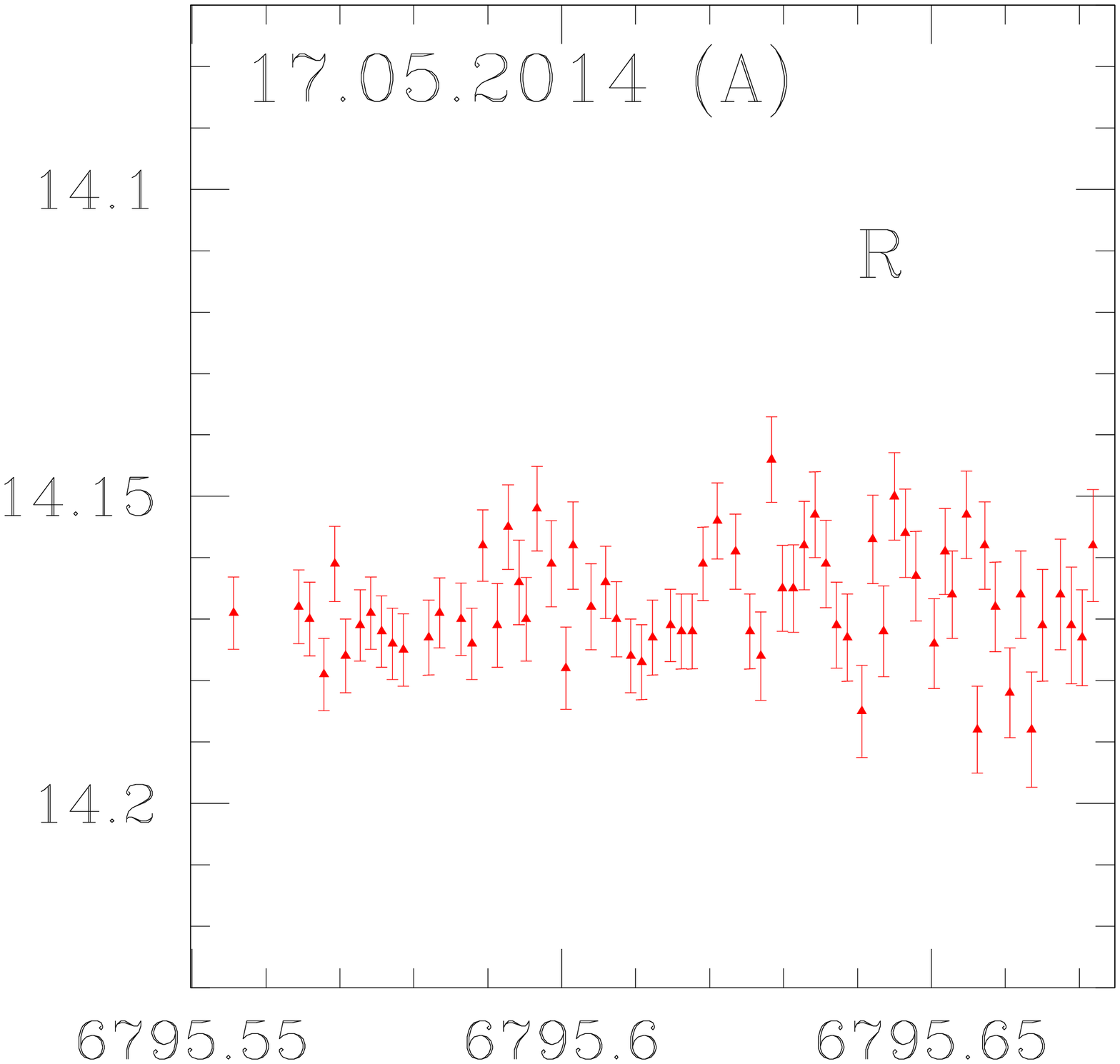,height=1.567in,width=1.59in,angle=0}
\epsfig{figure=  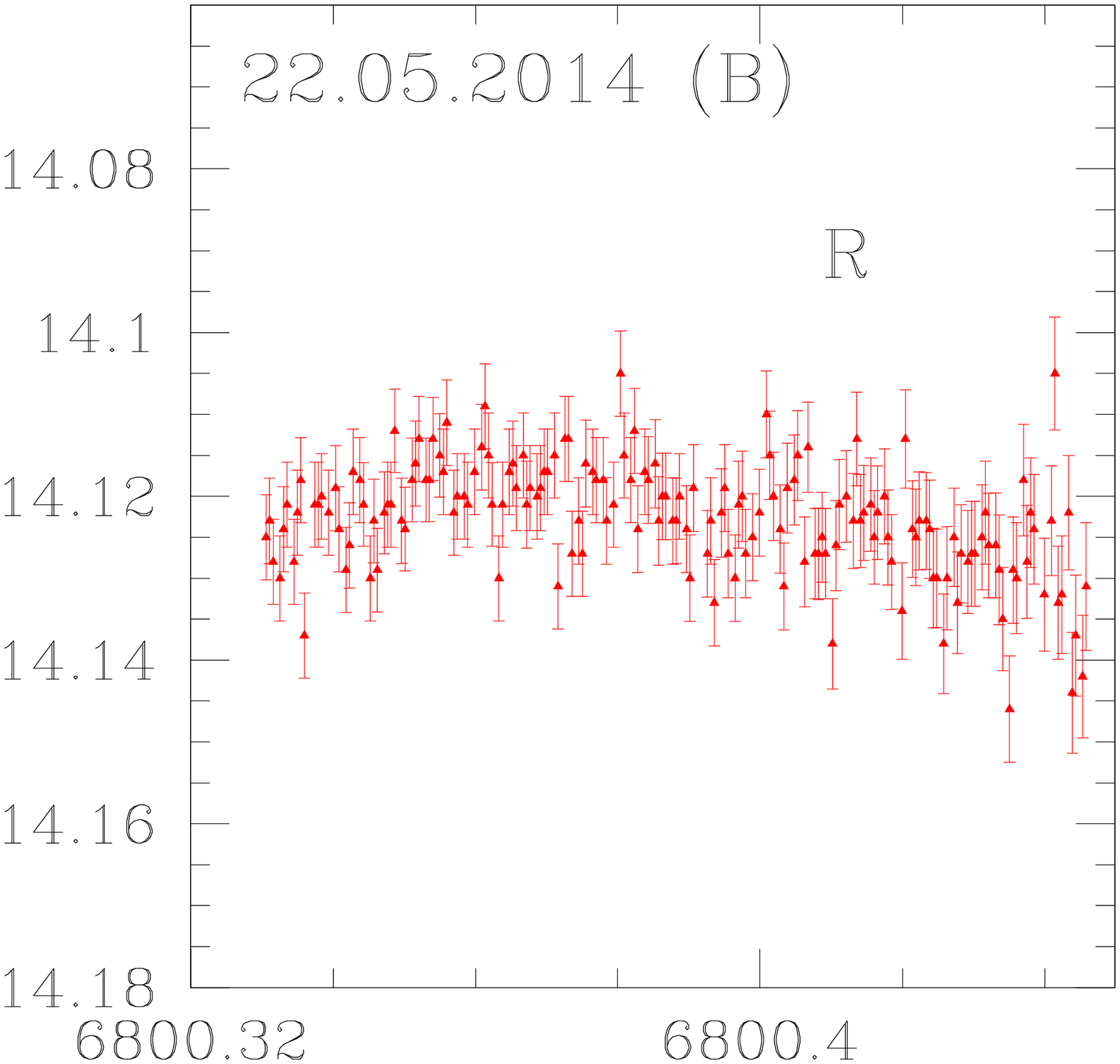,height=1.567in,width=1.59in,angle=0}
\epsfig{figure=  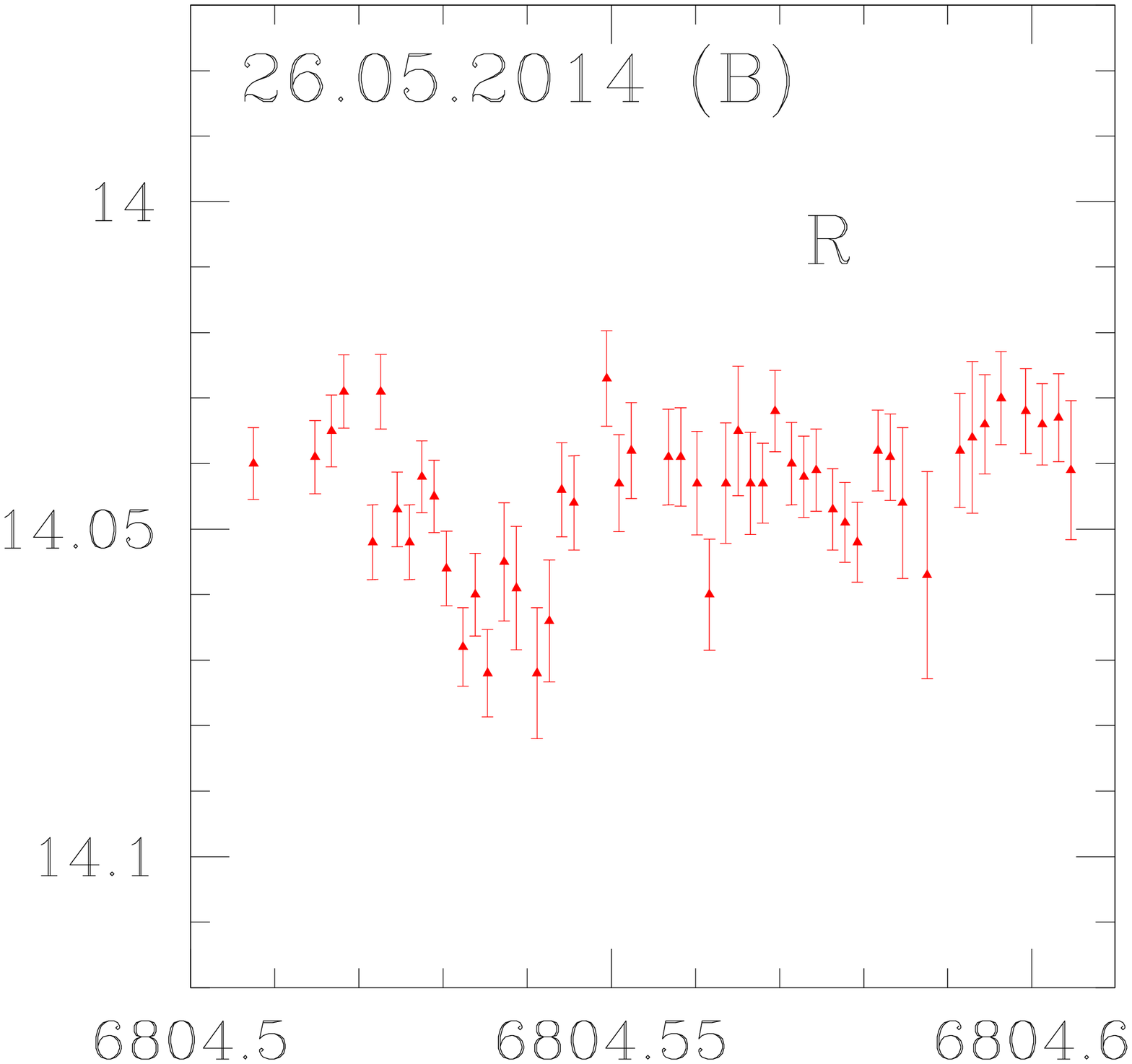,height=1.567in,width=1.59in,angle=0}
  \caption{Light curves for PG 1553+113.
  In each panel, the axes are the JD (+24590000) and R-band magnitude; the
observation date and the telescope used are indicated.}
\end{figure*}

\subsubsection{\bf $\chi^{2}$-test}

To examine the presence or absence of IDV we also performed $\chi^{2}$-test which is defined as (e.g.\ Agarwal \& Gupta 2015):

\begin{equation}
\chi^2 = \sum_{i=1}^N \frac{(V_i - \overline{V})^2}{\sigma_i^2},
\end{equation}
where, $\overline{V}$ is the mean magnitude, and the $i$th observation yields a magnitude $V_i$
with a corresponding standard error $\sigma_i$, which is due to photon noise from the source and sky, CCD read-out and other non-systematic
error sources. 
Exact quantification of such errors by the IRAF reduction package is impractical and it has been found that
theoretical errors are smaller than the real
errors by a factor of 1.3--1.75 (e.g., Gopal-Krishna et al.\ 2003) which for our data is $\sim$1.5, on average. 
So the errors obtained after data reduction 
should be multiplied by this factor to get better estimates of the real photometric errors.
This statistic is then compared with a critical value $\chi_{\alpha,\nu}^2$ where $\alpha$ is again the significance level  as the in case of
the F-test while $\nu = N -1$ is the number of degrees of freedom; $\chi^2 > \chi_{\alpha,\nu}^2$ implies the presence of variability.

\noindent

\subsubsection{\bf Percentage amplitude variation}

The percentage variation on a given night is calculated by using the variability amplitude parameter $A$,
introduced by Heidt \& Wagner (1996), and defined as
\begin{eqnarray}
A = 100\times \sqrt{{(A_{max}-A_{min}})^2 - 2\sigma^2}(\%) .
\end{eqnarray}
Here,  $A_{max}$ and $A_{min}$ are the maximum and minimum values in the calibrated LCs of the blazar, and $\sigma$
is the average measurement error.

\section{\bf  Results}

\subsection{PKS 1510$-$089}

PKS 1510$-$089 ($\alpha_{2000.0}$ = 15h 12m 50.53s, $\delta_{2000.0}$ = $-09^{\circ} 05^{'} 59^{\prime \prime}$)
is a FSRQ at a redshift of $z = 0.361$ (Thompson, Djorgovski, \& de Carvalho 1990) and is among the highly polarized AGN.
Lu (1972) first reported significant optical flux variations over a time span of $\sim$5 yr.
During a 1948 outburst, PKS 1510--089 showed an extremely large variation of $\Delta$B = 5.4 mag and later faded by $\sim$ 2.2 mag
within 9 days (Liller \& Liller 1975). On IDV timescales some very strong variations have been reported in optical bands: $\Delta$R of 0.65 mag
in 13 min (Xie et al.\ 2001) and of 2.0 mag in 42 min (Dai et al.\ 2001), while in V band a change of 1.68 mag in 60 min has been reported (Xie et al.\ 2002).
It was detected by the EGRET instrument on-board the CGRO in the MeV–-GeV energy band (Hartman et al.\ 1992) as $\gamma$-rays.
The synchrotron emission peaks around IR frequencies and  IC component seems to dominate the $\gamma$-rays . A pronounced UV bump is clearly visible in this source
which can be attributed to the thermal emission from the AD around the central region (Malkan \& Moore 1986; Pian \& Treves 1993).
A rapid $\gamma$-ray flare was reported by D'Ammando et al.\ (2009) in 2008 March using the AGILE satellite, while it was monitored by the Whole Earth Blazar Telescope.
Later, during the high $\gamma$-ray state of this source in 2009 March, D'Ammando et al.\ (2011) reported detailed analysis of their
multifrequency monitoring of this FSRQ using AGILE. They also found noticeable spectral variations in the near-IR through optical bands and
also found thermal features in the optical/UV spectrum in the broad-band SED of this source.

\begin{figure}
\epsfig{figure= 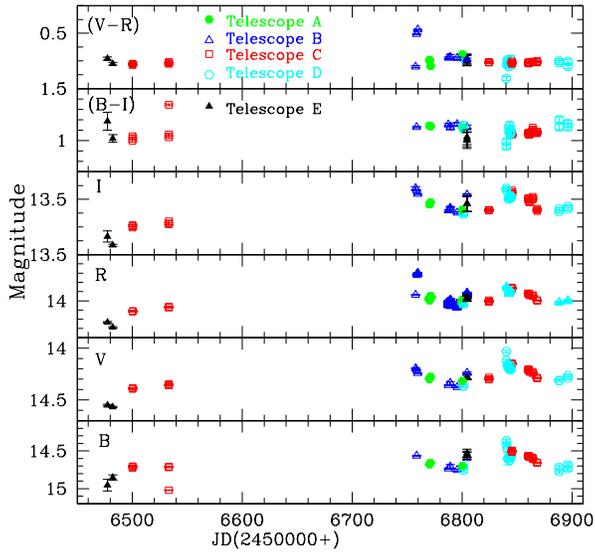,height=3.1in,width=3.1in,angle=0}
\caption{Long-term variability LCs and colour indices of PG 1553$+$113 in the B, V, R and I bands 
and (B-I) and (V-R) colours. Different symbols denote data from different observatories: filled green circles, telescope A; blue open triangles, telescope B;
red open squares, telescope C; cyan open squares, telescope D;   black closed triangles,telescope E.}
\end{figure}

\subsubsection{\bf Flux and colour variability}

The $\sim$ 4 months LCs for the blazar PKS 1510-089 in B, V, R, and I passbands are shown in Figure 1 along with the colour indices (B-I) and (V-R).
Genuine STV was present, but this source exhibited no significant colour variation during the observation span.
To provide a better view of the nature of the variability of this source around MJD 6840--6846, we have plotted this same time span on
an expanded scale in Figure 2.

In the following analysis, we have corrected the calibrated magnitudes for galactic extinction following the extinction map of Schlegel et al.\
(1998) using the NED extinction calculator\footnote{\tt http://ned.ipac.caltech.edu} with values in each filter as:
A$_{B}$ = 0.363 mag, A$_{V}$ = 0.275 mag, A$_{R}$ = 0.217 mag and A$_{I}$ = 0.151 mag (Cardelli et al.\ 1989; Bessell, Castelli, \& Plez 1998).
The bottom panel of Figure 1 represents short-term/long-term variability (STV/LTV) LC of PKS 1510$-$089 in the B passband which includes data
from 6 nights using the 1.3m RC telescope in Greece. We monitored the source in he B filter for STV/LTV studies from JD 2456824.37 to JD 2456868.27.
The maximum B band magnitude attained by the source was 16.77 on JD 2456824.37 while the brightest level of B = 16.41 was reached on JD 2456861.27,
which is $\sim$ 1.39 magnitude brighter than the faintest level of B = 17.8 as reported by Liller \& Liller (1975),  thus indicating that the source
is not in a  faint state and could possibly be in a post-outburst state.
The corresponding STV/LTV LCs in V, R, and I passbands are displayed in the second, third and fourth panels (from the bottom) of Figure  1 using data sets from 
the 1.3m RC telescope in Greece and the 60 cm Cassegrain telescope in Bulgaria (telescope C \& D) covering a time span between JD 2456778.37 to
JD 2456888.28.
The details of magnitude changes during the whole observation period are listed in Table 6 where column 1 is the source name, column 2 is the filter
used for observation, Column 3 tells the faintest magnitude attained by the target and its corresponding JD is given in column 4, while the
brightest value is given in column 5 , the time corresponding to it is given in column 6 and the last column gives the total magnitude range.
During the entire monitoring period, the overall magnitude variations were $\Delta$B = 0.36, $\Delta$V = 0.85, $\Delta$R = 0.75, and $\Delta$I = 0.83.

The (B-I) and (V-R) colour variations as a function of JD  are shown in the top two panels of Figure 1.
The maximum variation in (B-I) during our observation span was found to be a modest 0.29 mag, between 1.13 mag on JD 245684.38 and 1.42 mag on
JD 2456861.27 while that in (V-R) was
0.64 mag, between 0.17 mag on JD 2456844.35 and 0.81 mag on JD 2456888.28.
The mean magnitudes in B, V, R, and I are 16.60, 16.25, 15.86, and 15.26 mag,
respectively, while the average colour indices are (B-I) = 1.29 and (V-R) = 0.40.

\begin{table}
\caption{ Results of STV/LTV studies for magnitude changes in each band.  }
\textwidth=6.0in
\textheight=9.0in
\vspace*{0.2in}
\noindent
\begin{tabular}{p{1.5cm}p{0.3cm}p{0.5cm}p{1cm}p{0.6cm}p{1cm}p{0.4cm}} \hline

Source      & Band & Faintest  & ~~~JD  & Brightest & ~~~JD  & $\Delta m$\\ 
            &      & ~~Mag       & ~~~(Min) &~~Mag  & ~~~(Max) &           \\ \hline 

PKS           & B    & 16.77  & 2456824.4 & 16.41 & 2456861.3 & 0.36 \\
 1510$-$089     & V    & 16.69  & 2456888.3 & 15.84 & 2456799.4 & 0.85 \\
              & R    & 16.16  & 2456824.4 & 15.41 & 2456799.4 & 0.75 \\
              & I    & 15.58  & 2456824.4 & 14.75 & 2456799.4 & 0.83 \\
PG            & B    & 15.21  & 2456533.3 & 14.55 & 2456840.4 & 0.66 \\
  1553$+$113    & V    & 14.71  & 2456482.3 & 14.17 & 2456840.4 & 0.54 \\
              & R    & 14.40  & 2456482.3 & 13.81 & 2456759.7 & 0.59 \\
              & I    & 13.91  & 2456482.3 & 13.50 & 2456757.7 & 0.41 \\
Mrk 501       & B    & 15.06  & 2456487.4 & 14.65 & 2456806.4 & 0.41 \\
              & V    & 14.29  & 2456482.4 & 13.94 & 2456806.4 & 0.35 \\
              & R    & 13.90  & 2456482.4 & 13.55 & 2456949.3 & 0.35 \\
              & I    & 13.29  & 2456901.3 & 12.58 & 2456801.4 & 0.71 \\
 \hline
\end{tabular}  
\end{table}

\subsection{PG 1553$+$113 }
PG 1553$+$113 ($\alpha_{2000.0}$ = 15h 55m 43.04s, $\delta_{2000.0}$ = $+11^{\circ} 11^{'} 24.4^{\prime \prime}$)
was discovered in the Palomar-Green survey
of ultraviolet-excess objects as a 15.5 magnitude blue stellar
object (Green, Schmidt \& Liebert 1986). It is a bright optical source with R band magnitude varying from $\sim$ 13 to $\sim$ 15.5
(Miller at al.\ 1988) while is hads a mean V-band magnitude around 14 (Falomo \& Treves 1990; Osterman et al. 2006)
Due to its featureless spectra, this object was suggested to be a BL Lacertae object (Miller \& Green 1983)
and its redshift determination has always been a challenge. Based on low resolution UV spectra, Miller \& Green (1983)
estimated its redshift $\sim$ 0.37. Based on the detection of strong Ly$\alpha$ + O VI absorbers, Danforth et al.\ (2010) proposed
$z > 0.395$ while $z \leq 0.58$ from statistical arguments. Recently, Kapanadze (2013) found that the upper limit to its redshift should be smaller
than that proposed by Danforth et al.\ (2010). 
Its  classification as a HSP was determined from its SED  (Falomo \& Treves 1990; Donato, Sambruna, \& Gliozzi 2005).
Falomo et al.\ (1994) found its optical spectral index to be constant ($\alpha$ $\sim$ -1) using  observations between 1986 and 1991 while
it underwent a   variation of $\Delta$V = 1.4. PG 1553$+$113 has been observed through entire EM spectra from radio through
very high energy $\gamma$-rays  up to 1 TeV (Aharonian et al.\ 2006; Albert et al.\ 2007b). Its
log(F$_{2KeV}$ /F$_{5GHz}$ )values range from −4.99 to −3.88, where F$_{2KeV}$ is its 2 keV X-ray flux, while F$_{5GHz}$ is the
radio flux at 5 GHz (Osterman et al.\ 2006; Rector et al.\ 2003).

\subsubsection{\bf Flux and colour variability}

We monitored PG 1553+113 in the R passband on 7 nights for a span of $\sim$ 4 hours on each night
to investigate flux variability properties on intra-day timescales.
The IDV plots are displayed in Figure 3 where observation date and telescope used are mentioned within the plot itself.
The  observation log is given in Table 3.
To claim the presence or absence of intra-day variability we applied the F- and $\chi^{2}$- statisitical tests, the results of which are presented
in Table 5.
The blazar is said to be variable (V) if the variability conditions for both tests are satisfied for the 0.999 level, while it is marked
probably variable (PV) if conditions for either of the two tests are followed at the 0.99 level;
while the it is marked non-variable (NV) if none of
these conditions are met.
We detected strong microvariability on 1 night while it  was found to be PV on 5 nights and was clearly NV on only a single night.

The 1.2 year LCs in B, V, R, and I along with (B-I) and (V-R) colour variations for the blazar PKS 1553$+$113 are shown in Figure 4.
To study optical properties
on short/long timescales we investigated the target in B, V, R, and I filters observed using all 5 telescopes
on 29 nights during the period between JD 2456477.5 and JD 2456896.5.
We have corrected the calibrated magnitudes for galactic extinction in each filter with
A$_{B}$ = 0.188 mag, A$_{V}$ = 0.142 mag, A$_{R}$ = 0.113 mag and A$_{I}$ = 0.078 mag.
The details of the magnitude changes during the whole observation span for each band are listed in Table 6.
Significant STV was found with moderate colour variations. The minimum R band magnitude attained by our target was of R = 13.81
on JD = 2456759.7 which is
just 0.31 mag fainter than the brightest magnitude of R = 13.5 mag as observed earlier by Osterman et al.\ (2006), when the source was in a flaring state.
So it is fair to say that we have also observed this blazar in its flaring state.
During our observation time span, the overall magnitude variations were $\Delta$B = 0.66, $\Delta$V = 0.54, $\Delta$R = 0.59, and $\Delta$I = 0.41.

The values of the (B-I) and (V-R) colour indices as a function of JD  are shown in top two panels of Figure 4.
Plots indicate moderate colour variation.
The maximum variation noticed in the source for (V-R) during the entire LC was found to be 0.35, between 0.20 mag
on JD 2456840.37 and 0.55 mag on JD 2456759.70, while that in (B-I) was
1.39 between its colour range of 0.06 mag on JD 2456824.4 and 1.45 mag on JD 2456533.26.
Mean magnitudes in B, V, R, and I are 14.82, 14.43, 14.07, and 13.63,
while the average colour indices are (B-I) = 1.20 and (V-R) = 0.34.

\begin{figure}
\epsfig{figure= 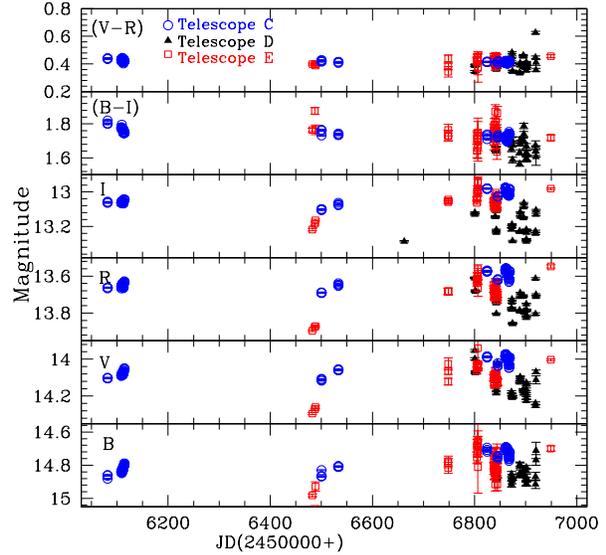,height=3.1in,width=3.1in,angle=0}
\caption{Long-term variability LCs and colour indices of Mrk 501 in the B, V, R and I bands 
and (B-I) and (V-R) colours. Different symbols denote data from different observatories: blue open circles, telescope C; black filled triangles, telescope D; and
red open squares, telescope E.}
\end{figure}

\begin{figure}
\epsfig{figure= 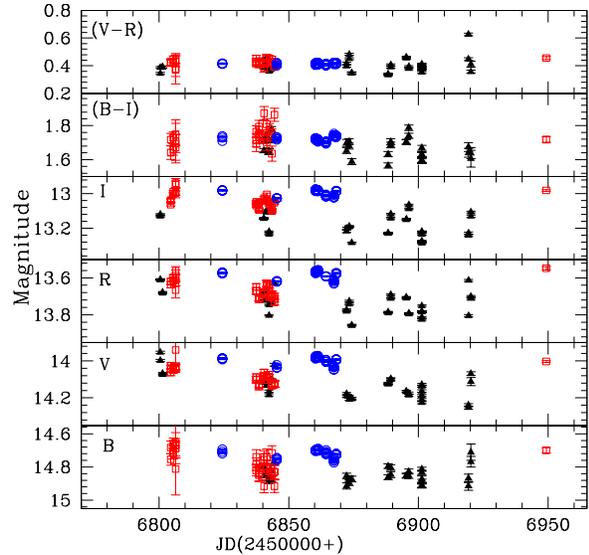,height=3.1in,width=3.1in,angle=0}
\caption{ Data from Figure 5 expanded  around MJD 6800-7000 to better show variability in that period.  Symbols as in Fig. 5.  }
\end{figure}

\subsection{Mrk 501}
Mrk 501 ($\alpha_{2000.0}$ = 16h 53m 52.13s, $\delta_{2000.0}$ = $+39^{\circ} 45^{'} 36.2^{\prime \prime}$)
is the second closest BL Lacertae object after Mrk 421, with $z = 0.034$, and is thus  an object of interest across the whole EM spectrum
(Fan \& Lin 1999; Kataoka et al.\ 1999; Sambruna 2000; Xue \& Cui 2005; Gliozzi et al.\ 2006).
The near-IR--optical spectrum of this HBL shows a strong host galaxy signature with host brightness of $\simeq$ 11.92 in R band
(Nilsson et al.\ 2007).
The SED of Mrk 501 displays a double-humped structure with peaks occurring at keV and GeV/TeV energies. The physical mechanisms responsible
for the production of GeV/TeV hump are still a topic of debate (Ghisellini \& Madau 1996; Krawczynski et al.\ 2004; Cerruti et al.\ 2012).
In the optical regime, Heidt \& Wagner (1996)
reported a flux variation of $\sim$32\% in a time span of less than 2 weeks while Ghosh et al.\ (2000) reported variability in 7 out of 10 nights
during March \& June 1997.
Mrk 501 has been found to display rapid variability on few minutes timescales over the entire EM spectra (Albert et al.\ 2007a; Gupta et al.\ 2008a), 
which can be attributed to relativistically beamed radiation from jets, jet deceleration (Georganopoulos \& Kazanas 2003; Levinson 2007),
or other plasma mechanisms (e.g.\ Krishan \& Wiita 1994)

\begin{figure*}
\epsfig{figure= 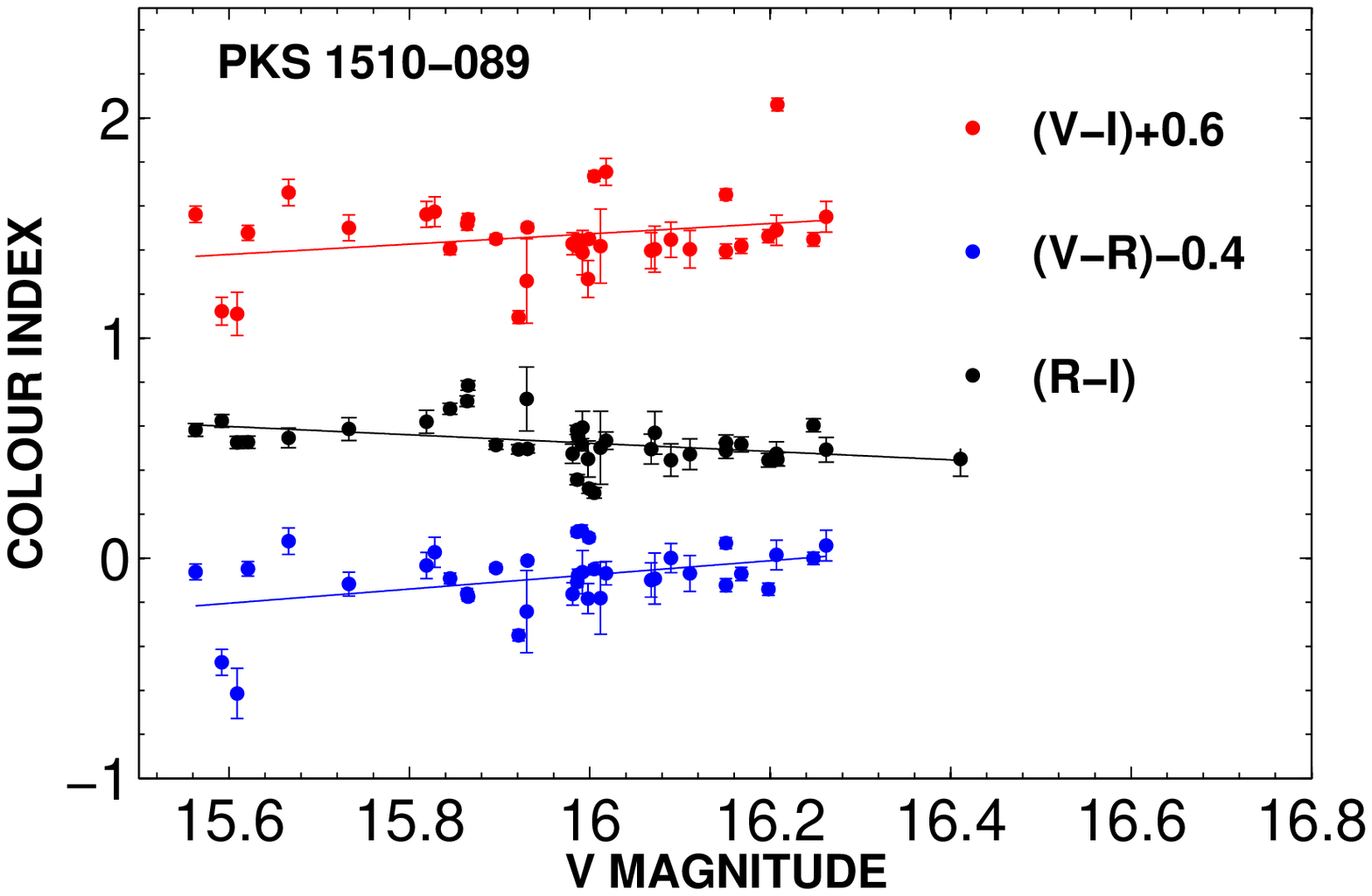,height=2.0in,width=2.3in,angle=0}
\epsfig{figure= 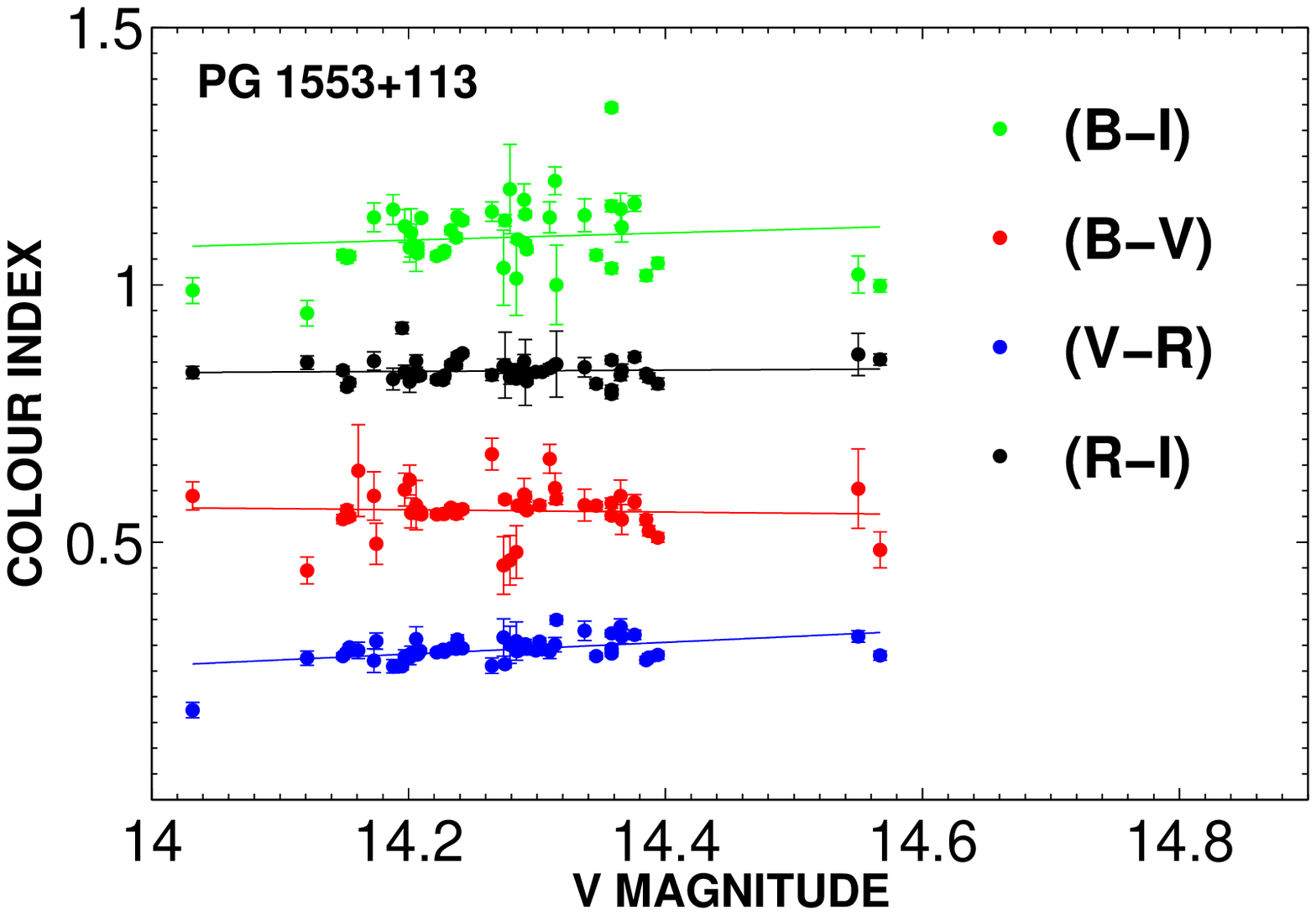,height=2.0in,width=2.3in,angle=0}
\epsfig{figure= 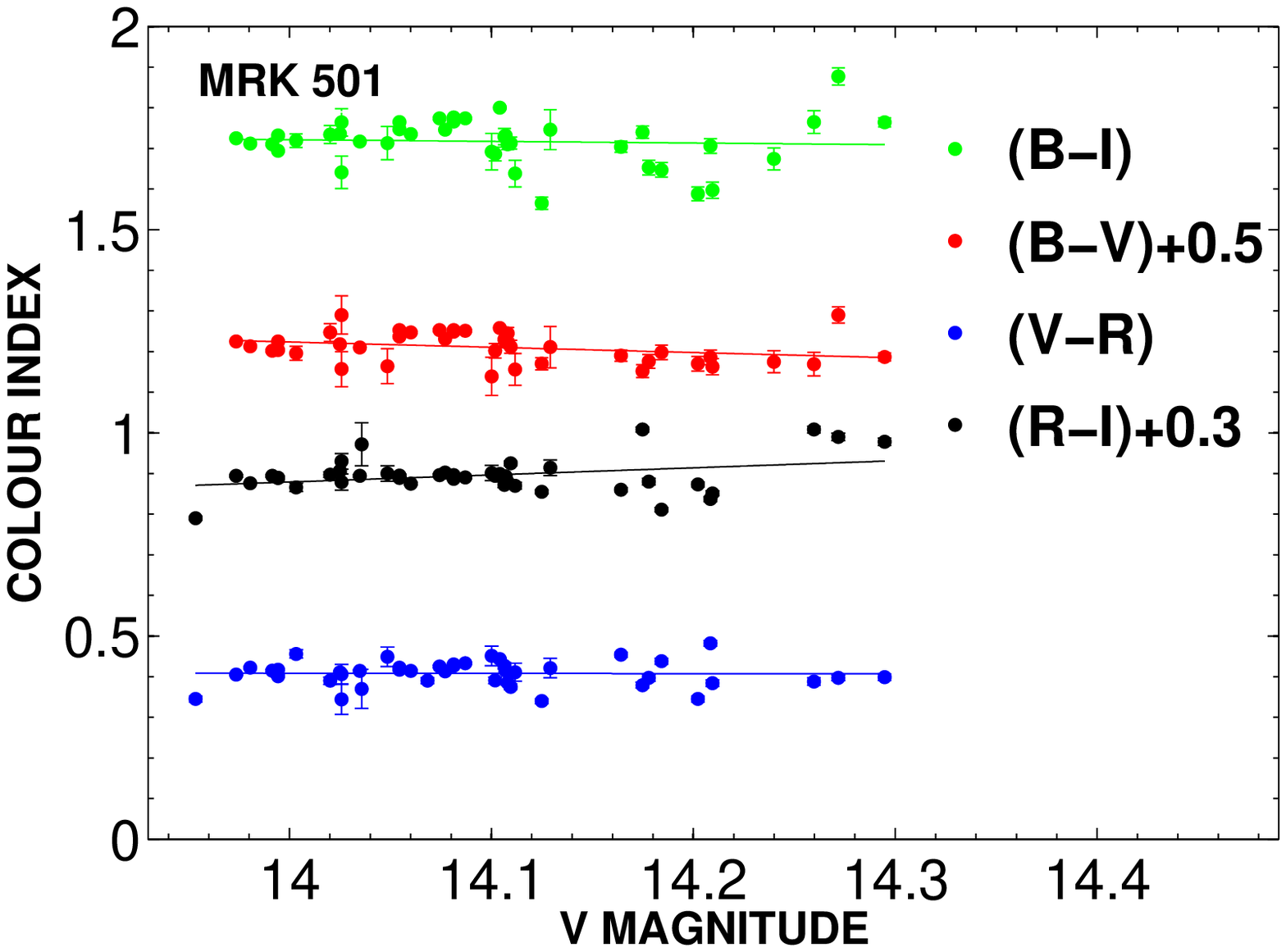,height=2.0in,width=2.3in,angle=0}
\caption{Colour magnitude plots on short timescales for all three blazars.  
The V magnitudes are given on the X-axis and the various labeled colour indices are plotted 
against them for each labeled  source.}
\end{figure*}

\subsubsection{\bf Flux and colour variability}

The photometric magnitudes extracted during our monitoring campaign are displayed in figure 5 which shows $\sim$ 2.3 year LC for the blazar MRK 501.
The LCs in B, V, R, and I passbands along with (B-I) and (V-R) colour variations are plotted in different panels of above figure.
Figure 6 shows the nature of variability in more detail around MJD 6800-7000, when this blazar displays large variability.

The target was observed for 48 nights between 2456081.5 and 2456920.4 using five telescopes whose details are given in Table 1.
Details about magnitude changes in each band during the whole monitoring period are given in Table 6.
In the following analysis, we have corrected the calibrated magnitudes for galactic extinction as:
A$_{B}$ = 0.069 mag, A$_{V}$ = 0.052 mag, A$_{R}$ = 0.041 mag and A$_{I}$ = 0.029 mag.
The flux from the nucleus of the HBL is contaminated
by the emissions of its host galaxy. So the observed magnitudes in each spectral band have also been corrected for the host galaxy contribution,
following the measurements of
Nilsson et al.\ (2007) to calculate the host galaxy contribution in R band which is then used to find the corresponding contributions for the
B, V, and I  bands (Fukugita et al.\ 1995).

During our observation time span, the overall magnitude variations were $\Delta$B = 0.41, $\Delta$V = 0.35, $\Delta$R = 0.35, and $\Delta$I = 0.71.
The mean magnitudes in B, V, R, and I are 14.80, 14.09, 13.68, and 13.07,
respectively.
The short-term and long-term LCs of Mrk 501 in (V-R) and (B-I) colours are shown in the top two panels of Figure 5.
The average colour indices are (B-I) = 1.72 and (V-R) = 0.41 
The maximum variation noticed in the source for the (V-R) colour index during the entire LC was found to be 0.29 (the range was between 0.33 mag
on JD 2456888.33 and 0.63 mag on JD 2456919.31), while that in (B-I) was
0.31 (this colour index ranged between 1.56 mag on JD 2456888.33 and 1.88 mag on JD 2456487.40).

\subsection{\bf Correlated variations between colour and magnitude?}

As variations in the optical flux of blazars are accompanied with spectral changes, studying the colour index--magnitude relationship
can be an useful tool to understand the origin of variability in blazars.
In this section, we investigate the colour--magnitude relationship for all  three blazars.
Since spectral variations follow any optical flux variations,  examining relationships between the corresponding variations in the colour indices
such as (B-V), (V-R), (R-I), (B-R) or (B-I) of the three targets with respect to the variation in their brightnesses would be very helpful in understanding the
variability characteristics in more detail.

\begin{table}
\caption{Color--magnitude dependencies and colour-magnitude correlation coefficients on short timescales for PKS 1510-089.  }
\textwidth=7.0in
\textheight=10.0in
\vspace*{0.2in}
\noindent

\begin{tabular}{ccccc} \hline 

Color Indices     &  $m_1^a$  &  $c_1^a$  &   $r_1^a$  & $p_1^a$    \\ \hline
(V-I)          & 0.234 & $-$2.571 & 0.250 & 0.130 \\ 
(V-R)          &  0.322 & $-$5.227 & 0.397  & 0.015  \\     
(R-I)          &  $-$0.189 & 3.554 & $-$0.374 & 0.021 \\
\hline
\end{tabular}\\
$^a$ $m_1 =$ slope and $c_1 =$ intercept of CI against V; \\
$r_1 =$ Pearson coefficient; $p_1 =$ null hypothesis probability \\
\end{table}

\begin{table}
\caption{Color--magnitude dependencies and colour-magnitude correlation coefficients on short timescales for PG 1553+113.  }
\textwidth=7.0in
\textheight=10.0in
\vspace*{0.2in}
\noindent

\begin{tabular}{ccccc} \hline 

Color Indices     &  $m_1^a$  &  $c_1^a$  &   $r_1^a$  & $p_1^a$    \\ \hline
(B-I)          &  0.070 & 0.093 & 0.102 & 0.498 \\
(B-V)          &  $-$0.022 & 0.876 & $-$0.048  & 0.744  \\     
(V-R)          &  0.114 & $-$1.336 & 0.440 & 0.0009 \\
(R-I)          &  0.012 & 0.660 & 0.054 & 0.704  \\
\hline
\end{tabular}\\
$^a$ $m_1 =$ slope and $c_1 =$ intercept of CI against V; \\
$r_1 =$ Pearson coefficient; $p_1 =$ null hypothesis probability \\
\end{table}

\begin{table}
\caption{Color--magnitude dependencies and colour-magnitude correlation coefficients on short timescales for Mrk 501.  }
\textwidth=7.0in
\textheight=10.0in
\vspace*{0.2in}
\noindent

\begin{tabular}{ccccc} \hline 

Color Indices     &  $m_1^a$  &  $c_1^a$  &   $r_1^a$  & $p_1^a$    \\ \hline
(B-I)          &  $-$0.040 & 2.282 & $-$0.057 & 0.722 \\
(B-V)          &  $-$0.130 & 3.042 & $-$0.287  & 0.069  \\     
(V-R)          &  $-$0.003 & 0.454 & $-$0.009 & 0.956 \\
(R-I)          &  0.174 & $-$1.559 & 0.329 & 0.033  \\ 
\hline
\end{tabular}\\
$^a$ $m_1 =$ slope and $c_1 =$ intercept of CI against V; \\
$r_1 =$ Pearson coefficient; $p_1 =$ null hypothesis probability \\
\end{table}

Colour--magnitude (CM) plots on few months timescales for the three sources are shown in Figure 7.
We have fitted straight lines (CI =$m$V + $c$) to colour index, CI against V magnitude plots for each source.  The fitted values for
the slope, $m$, and intercept $c$, are listed in Tables 7, 8 and 9 for PKS 1510-089, PG 1553+113 and Mrk 501, respectively,
along with the linear Pearson correlation coefficients, $r_1$ and the corresponding null 
hypothesis probabilities, $p_1$. 
A positive slope here means positive correlation between the CI and the brightness of the source. Here we consider
the result significant only if the null 
hypothesis probability is $p_1 \le 0.01$ and Pearson correlation coefficient, $r \ge 0.5$.
A significant positive correlation in these plots would physically imply that the source follows a bluer-when-brighter (BWB; or redder when fainter) trend, i.e.,
the source tends to be bluer when its brightness increases, while a negative slope would indicate an opposite correlation between the source
magnitude and CI,  indicating that the source exhibits redder-when-brighter behaviour (RWB) behaviour.

The correlation analysis results between brightness of the source and the colour indices are also given in Tables 7--9. 
For these sources we did not find any strong correlation (Pearson correlation coefficient, $r > 0.5$)
between V band magnitude and colour indices on several months timescale.

The CM relationship at diverse timescales can help us understand the emission mechanisms responsible
for blazar variability and also help pin down the emitting regions.
Different CM relationship in BL Lacs and FSRQs could be due to the existence of two varying modes, i.e., larger flux variations with lesser spectral
changes or vice versa.
Raiteri et al.\ (2003) found at most weak correlations between the source magnitude and the colour indices for the BL Lacerate object
S5 0716+714, that are similar to our results.
Agarwal et al.\ (2016) also found no evidence for the source to display spectral changes with magnitude on either of the timescales even when the
BL Lac S5 0716+714 was rapidly variable.
The CM relation in blazars has been found to vary among their outburst state, active state, and the faint state (Sasada et al.\ 2010).
Optical emission from blazars is generally a combination of jet and AD radiation and usually dominated by that from the
relativistic jets.   However, FSRQs are usually low frequency synchrotron peaked sources, thus having a substantial contribution to the blue-UV continuum
arising from thermal contamination from the AD which is expected to produce a slowly and weakly variable bluer emission component.
The position of the thermal blue bump and its strength in comparison to the jet emission also affect the spectral behaviour of FSRQs.
When relativistically beamed jet synchrotron emission dominates the AD emission, the BWB trend can be explained by acceleration of relativistic
particles or due to injection of fresh electrons having even harder energy distribution.
Sun et al.\ (2014) recently proposed a timescale dependent colour variation model, according to which a BWB trend in blazars is strongest for
timescales of $<$ 30 days and it eventually weakens with as timescale increases to above 100 days. Our results seem to complement this model.
Densely sampled and simultaneous multiband photometric observations will be helpful in understanding the CM relationship in greater detail and in better contraining models.

\section{Discussion and Conclusion}

Blazar variability studies  can help us examine the radiation mechanisms in more detail while also provide an insight on the location, size,
structure, and dynamics of the emitting regions (Ciprini et al.\ 2003). Most mechanisms for variability in RLAGNs are expected to 
arise within the relativistic jet whose emission is Doppler boosted. The only significant extrinsic variability in blazars is probably seen
in the radio band, where strongly frequency dependent interstellar scintillation is found to be dominant mechanism of variability at low radio frequencies.
Intrinsic processes operate across all EM bands, and include directly those causing
changes in the jet radio emission.  Some of these clearly arise from synchrotron and Compton losses which cause electron energy losses but
energy gains can come from the fresh injection of particles with
energy distribution higher than that of the previous ones.  Both of these processes  can co-exist but operate on different timescales.

The initial origin of variability could be located in AD based fluctuations propagating into the jets, or it could start with 
changes in the outgoing flow, usually related to shocks in the blazar jets. 
According
to the dominant shock-in-jet model, synchrotron and inverse Compton (IC) processes are the most common basic processes
(Hughes, Aller, \& Aller 2011).
When those intense emitting regions are found in the highly relativistic Doppler
boosted jets pointed at very small angles with respect to the LOS, variations are significantly amplified and frequency shifted owing to relativistic
beaming.
Changes in the jet geometry due to changing jet direction can cause variations in the bulk Doppler factor of the relativistic blobs traveling
along the jet (usually in range 10-30), which in turn can lead to blazar variability at realtively long time-scales
(Hovatta et al.\ 2009; Rani et al. 2011).
The key processes
responsible for blazar variability probably are due to a jet not being a stationary object which gives rise to various instabilities,
turbulence (e.g.\ Marscher 2014; Calafut \& Wiita 2015) and developing or decaying shocks.  These act on a range of timescales.
 If $\delta$ is the the Doppler factor, since $F_{\nu} \propto  \delta^{\sim 3}$,  an increase in $\delta$ causes an increase in flux
(Villata \& Raiteri 1999) and also in the observed frequency ($\nu \propto  \delta$).
During their low states, blazar variablity can be attributed to AD instabilities since thermal emission from the central region of blazars
can dominate over jet emission then.
Then the variability can be explained by orbits of hot spots on the AD, including eclipsing of the hot spot by parts of the disk between the individual spot and the
observer;  this aspect directly depends on the geometry of the AD and also on the viewing angle of the observer.
Even though optical band is a narrow part of the complete EM spectrum, it is critical in determinibg 
the presence of additional components other than synchrotron continuum, such as AD emission or host galaxy contribution.

The presence of both BWB and RWB trends in some blazars can be explained by superposition of both blue and red emission components where the redder one
is attributed to the synchrotron radiation from the relativistic jet while the blue component could come from the thermal emission from the AD.
The BWB trend may indicate that two components, one variable (with a flatter slope,
$\alpha_{var}$) ($f_{\nu} \propto \nu^{-\alpha}$) and another stable (with $\alpha_{const} > \alpha_{var}$), contribute to the overall
emission in the optical regime. It also could  be possibly explained with a one component synchrotron model if the more intense the
energy release, the higher the particles' frequency (Fiorucci, Ciprini, \& Tosti et al.\ 2004). Then the BWB trend could be explained if the
luminosity increase was due to injection of fresh electrons with an energy distribution harder than that of the previously cooled
ones (Kirk et al.\ 1998; Mastichiadis \& Kirk 2002). The BWB trend could also be due to Doppler factor variations in a spectrum slightly deviating
from a power law (Villata et al.\ 2004).  
If optical emission is combination of emission from both jet and AD, then as the jet brightens from a low state, the colour of the combined emission gets redder
as synchrotron emission from the relativistic jets is intrinsically redder than that of the AD, producing RWB behaviour. However, if even higher energy electrons are injected
 causing further brightening, then we get a bluer colour (BWB); this is usually dominant during an outburst state.
We found a weak    BWB  trend, i.e., a spectrum becoming flatter when the object is brighter, in two of the sources while the opposite was found dominant in one of them.

In this paper, we have performed multiband optical photometry for three TeV blazars namely: PKS 1510$-$089, PG 1553$+$113 and Mrk 501 between 2012 and
2014 in a total of 95 nights in B, V, R, and I passbands.  This allowed us to study flux and colour variability characteristics on diverse timescales.
During our 7 nights of observation for IDV for  PG 1553$+$113 we detected clear microvariability on a single night  using F-statistics and $\chi^{2}$-test (at 3$\sigma$ significance for both).
But if we include PV cases (at $2.6 \sigma$ for either tests) then we can say we found the source to be variable on intraday timescales on 6 out of these 7 nights.
Carini (1990) monitored about 20 blazars and found IDV in most of them. He also noticed that probability of detecting IDV increased to 80\% when
observed for more than 8 hours. Later, a sample of 34 BL Lacerate objects were observed by Heidt \& Wagner (1996) when they found that about
75\% of the sample displayed significant variations when observed for $<$ 6 hours.  Carini et al.\ (2007) found that less than 10\% of his blazar sample
  displayed variability on intraday timescales
when observed for only $\sim$ 4-5 hours. Thus, the chances of observing IDV in blazars is greatly improved when
observed for more than 6 hours.  The results in this study are consistent with these earlier observations. 
Clearly, IDV results can be further improved by more dense and lengthier observations.

We  searched for flux and colour variability on few months to few years timescales and found significant flux variations on these 
timescales with moderate colour variations for all three TeV blazars.
We  also studied the correlation between the V magnitude of the source and corresponding variations in the (B-I), (B-V), (R-I),
and (V-R) colour indices. Our observations did not
reveal the presence of significant correlated spectral variability in these targets on short timescales.
Variability studies of larger blazar samples on minutes to years timescales are extremely important since they
can provide information on numerous blazar parameters.

\noindent
\section*{Acknowledgments}
We thank the referee for very useful comments which helped us to improve the manuscript.
This research was partially supported by Scientific Research Fund of the
Bulgarian Ministry of Education and Sciences under grant DO 02-137
(BIn-13/09).
GD and OV gratefully acknowledge the observing
grant support from the Institute of Astronomy and Rozhen National
Astronomical Observatory, Bulgaria Academy of Sciences. This
work is a part of the Projects No 176011 (Dynamics and kinematics
of celestial bodies and systems), No. 176004 (Stellar physics)
and No. 176021 (Visible and invisible matter in nearby galaxies:
theory and observations) supported by the Ministry of Education,
Science and Technological Development of the Republic of Serbia.

\label{lastpage}

\begin{thebibliography}{}

\bibitem[\protect\citeauthoryear{Abdo et al.}{2010}]{2010ApJ...716...30A} 
Abdo A.~A., et al., 2010, ApJ, 716, 30 

\bibitem[\protect\citeauthoryear{Agarwal 
\& Gupta}{2015}]{2015MNRAS.450..541A} Agarwal A., Gupta A.~C., 2015, MNRAS, 450, 541 

\bibitem[\protect\citeauthoryear{Agarwal et 
al.}{2015}]{2015MNRAS.451.3882A} Agarwal A., et al., 2015, MNRAS, 451, 3882 


\bibitem[\protect\citeauthoryear{Agarwal et 
al.}{2016}]{2016MNRAS.455..680A} Agarwal A., et al., 2016, MNRAS, 455, 680 

\bibitem[\protect\citeauthoryear{Aharonian et 
al.}{2006}]{2006A&A...448L..43A} Aharonian F., et al., 2006, A\&A, 448, L43

\bibitem[\protect\citeauthoryear{Albert et al.}{2007a}]{2007ApJ...669.1143A} 
Albert J., et al., 2007a, ApJ, 669, 1143 

\bibitem[\protect\citeauthoryear{Albert et al.}{2007b}]{2007ApJ...654L.119A} 
Albert J., et al., 2007b, ApJ, 654, L119 

\bibitem[\protect\citeauthoryear{Barkov et al.}{2012}]{2012ApJ...749..119B} 
Barkov M.~V., Aharonian F.~A., Bogovalov S.~V., Kelner S.~R., Khangulyan 
D., 2012, ApJ, 749, 119 

\bibitem[\protect\citeauthoryear{Begelman, Fabian, 
\& Rees}{2008}]{2008MNRAS.384L..19B} Begelman M.~C., Fabian A.~C., Rees M.~J., 2008, MNRAS, 384, L19

\bibitem[\protect\citeauthoryear{Bessell, Castelli, 
\& Plez}{1998}]{1998A&A...333..231B} Bessell M.~S., Castelli F., Plez B., 1998, A\&A, 333, 231 

\bibitem[\protect\citeauthoryear{Calafut 
\& Wiita}{2015}]{2015JApA...36..255C} Calafut V., Wiita P.~J., 2015, JApA, 36, 255 

\bibitem[\protect\citeauthoryear{Cardelli, Clayton, 
\& Mathis}{1989}]{1989ApJ...345..245C} Cardelli J.~A., Clayton G.~C., Mathis J.~S., 1989, ApJ, 345, 245

\bibitem[\protect\citeauthoryear{Carini}{1990}]{1990PhDT.......263C} Carini 
M.~T., 1990, PhDT, 

\bibitem[\protect\citeauthoryear{Carini et al.}{2007}]{2007AJ....133..303C} 
Carini M.~T., Noble J.~C., Taylor R., Culler R., 2007, AJ, 133, 303 

\bibitem[\protect\citeauthoryear{Cerruti et 
al.}{2012}]{2012AIPC.1505..635C} Cerruti M., Zech A., Boisson C., Inoue S., 
2012, AIPC, 1505, 635 


\bibitem[\protect\citeauthoryear{Ciprini et 
al.}{2003}]{2003A&A...400..487C} Ciprini S., Tosti G., Raiteri C.~M., Villata M., Ibrahimov M.~A., Nucciarelli G., Lanteri L., 2003, A\&A, 400, 487 


\bibitem[\protect\citeauthoryear{Clements, Jenks, 
\& Torres}{2003}]{2003AJ....126...37C} Clements S.~D., Jenks A., Torres Y., 2003, AJ, 126, 37

\bibitem[\protect\citeauthoryear{D'Ammando et 
al.}{2009}]{2009A&A...508..181D} D'Ammando F., et al., 2009, A\&A, 508, 181

\bibitem[\protect\citeauthoryear{D'Ammando et 
al.}{2011}]{2011A&A...529A.145D} D'Ammando F., et al., 2011, A\&A, 529, A145 

\bibitem[\protect\citeauthoryear{Dai et al.}{2001}]{2001AJ....122.2901D} 
Dai B.~Z., Xie G.~Z., Li K.~H., Zhou S.~B., Liu W.~W., Jiang Z.~J., 2001, 
AJ, 122, 2901

\bibitem[\protect\citeauthoryear{Danforth et 
al.}{2010}]{2010ApJ...720..976D} Danforth C.~W., Keeney B.~A., Stocke 
J.~T., Shull J.~M., Yao Y., 2010, ApJ, 720, 976 

\bibitem[\protect\citeauthoryear{Donato, Sambruna, 
\& Gliozzi}{2005}]{2005A&A...433.1163D} Donato D., Sambruna R.~M., Gliozzi M., 2005, A\&A, 433, 1163

\bibitem[\protect\citeauthoryear{Falomo 
\& Treves}{1990}]{1990PASP..102.1120F} Falomo R., Treves A., 1990, PASP, 102, 1120 

\bibitem[\protect\citeauthoryear{Falomo, Scarpa, 
\& Bersanelli}{1994}]{1994ApJS...93..125F} Falomo R., Scarpa R., Bersanelli M., 1994, ApJS, 93, 12

\bibitem[\protect\citeauthoryear{Fan 
\& Lin}{1999}]{1999ApJS..121..131F} Fan J.~H., Lin R.~G., 1999, ApJS, 121, 131 

\bibitem[\protect\citeauthoryear{Fiorucci, Ciprini, 
\& Tosti}{2004}]{2004A&A...419...25F} Fiorucci M., Ciprini S., Tosti G., 2004, A\&A, 419, 25 

\bibitem[\protect\citeauthoryear{Fossati et 
al.}{1997}]{1997MNRAS.289..136F} Fossati G., Celotti A., Ghisellini G., 
Maraschi L., 1997, MNRAS, 289, 136

\bibitem[\protect\citeauthoryear{Fukugita, Shimasaku, 
\& Ichikawa}{1995}]{1995PASP..107..945F} Fukugita M., Shimasaku K., Ichikawa T., 1995, PASP, 107, 945

\bibitem[\protect\citeauthoryear{Gaur, Gupta, 
\& Wiita}{2012}]{2012AJ....143...23G} Gaur H., Gupta A.~C., Wiita P.~J., 2012, AJ, 143, 23 

\bibitem[\protect\citeauthoryear{Georganopoulos 
\& Kazanas}{2003}]{2003ApJ...594L..27G} Georganopoulos M., Kazanas D., 2003, ApJ, 594, L27 

\bibitem[\protect\citeauthoryear{Ghisellini 
\& Madau}{1996}]{1996MNRAS.280...67G} Ghisellini G., Madau P., 1996, MNRAS, 280, 67 

\bibitem[\protect\citeauthoryear{Ghosh et al.}{2000}]{2000ApJS..127...11G} 
Ghosh K.~K., Ramsey B.~D., Sadun A.~C., Soundararajaperumal S., 2000, ApJS, 
127, 11 

\bibitem[\protect\citeauthoryear{Gliozzi et 
al.}{2006}]{2006ApJ...646...61G} Gliozzi M., Sambruna R.~M., Jung I., 
Krawczynski H., Horan D., Tavecchio F., 2006, ApJ, 646, 61 

\bibitem[\protect\citeauthoryear{Gopal-Krishna et 
al.}{2003}]{2003ApJ...586L..25G} Gopal-Krishna, Stalin C.~S., Sagar R., 
Wiita P.~J., 2003, ApJ, 586, L25

\bibitem[\protect\citeauthoryear{Green, Schmidt, 
\& Liebert}{1986}]{1986ApJS...61..305G} Green R.~F., Schmidt M., Liebert J., 1986, ApJS, 61, 305 

\bibitem[\protect\citeauthoryear{Gupta et 
al.}{2004}]{2004A&A...422..505G} Gupta A.~C., Banerjee D.~P.~K., Ashok N.~M., Joshi U.~C., 2004, A\&A, 422, 505 

\bibitem[\protect\citeauthoryear{Gupta 
\& Joshi}{2005}]{2005A&A...440..855G} Gupta A.~C., Joshi U.~C., 2005, A\&A, 440, 855

\bibitem[\protect\citeauthoryear{Gupta et al.}{2008a}]{2008NewA...13..375G} 
Gupta A.~C., Deng W.~G., Joshi U.~C., Bai J.~M., Lee M.~G., 2008a, NewA, 13, 
375 

\bibitem[\protect\citeauthoryear{Gupta et al.}{2008b}]{2008AJ....135.1384G}
Gupta A.~C., Fan J.~H., Bai J.~M., Wagner S.~J., 2008b, AJ, 135, 1384 

\bibitem[\protect\citeauthoryear{Hartman et 
al.}{1992}]{1992ApJ...385L...1H} Hartman R.~C., et al., 1992, ApJ, 385, L1

\bibitem[\protect\citeauthoryear{Heeschen et 
al.}{1987}]{1987AJ.....94.1493H} Heeschen D.~S., Krichbaum T., Schalinski 
C.~J., Witzel A., 1987, AJ, 94, 1493 

\bibitem[\protect\citeauthoryear{Heidt
\& Wagner}{1996}]{1996A&A...305...42H} Heidt J., Wagner S.~J., 1996, A\&A, 305, 

\bibitem[\protect\citeauthoryear{Holder}{2012}]{2012APh....39...61H} Holder 
J., 2012, APh, 39, 61 

\bibitem[\protect\citeauthoryear{Holder}{2014}]{2014BrJPh..44..450H} Holder 
J., 2014, BrJPh, 44, 450 

\bibitem[\protect\citeauthoryear{Hovatta et 
al.}{2009}]{2009A&A...494..527H} Hovatta T., Valtaoja E., Tornikoski M., L{\"a}hteenm{\"a}ki A., 2009, A\&A, 494, 527

\bibitem[\protect\citeauthoryear{Hughes, Aller, 
\& Aller}{2011}]{2011ApJ...735...81H} Hughes P.~A., Aller M.~F., Aller H.~D., 2011, ApJ, 735, 81 

\bibitem[\protect\citeauthoryear{Jannuzi, Smith, 
\& Elston}{1994}]{1994ApJ...428..130J} Jannuzi B.~T., Smith P.~S., Elston R., 1994, ApJ, 428, 130

\bibitem[\protect\citeauthoryear{Kapanadze}{2013}]{2013AJ....145...31K} 
Kapanadze B.~Z., 2013, AJ, 145, 31 


\bibitem[\protect\citeauthoryear{Kataoka et 
al.}{1999}]{1999ApJ...514..138K} Kataoka J., et al., 1999, ApJ, 514, 138 

\bibitem[\protect\citeauthoryear{Kinman}{1975}]{1975IAUS...67..573K} Kinman 
T.~D., 1975, IAUS, 67, 573

\bibitem[\protect\citeauthoryear{Kirk et al.}{1998}]{1998LPI....29.1752K} 
Kirk R.~L., et al., 1998, LPI, 29, 1752 

\bibitem[\protect\citeauthoryear{Krishan 
\& Wiita}{1994}]{1994ApJ...423..172K} Krishan V., Wiita P.~J., 1994, ApJ, 423, 172

\bibitem[\protect\citeauthoryear{Krawczynski et 
al.}{2004}]{2004ApJ...601..151K} Krawczynski H., et al., 2004, ApJ, 601, 
151

\bibitem[\protect\citeauthoryear{Levinson}{2007}]{2007ApJ...671L..29L} 
Levinson A., 2007, ApJ, 671, L29 

\bibitem[\protect\citeauthoryear{Liller 
\& Liller}{1975}]{1975ApJ...199L.133L} Liller M.~H., Liller W., 1975, ApJ, 199, L133

\bibitem[\protect\citeauthoryear{Lu}{1972}]{1972AJ.....77..829L} Lu P.~K., 
1972, AJ, 77, 829 

\bibitem[\protect\citeauthoryear{Malkan 
\& Moore}{1986}]{1986ApJ...300..216M} Malkan M.~A., Moore R.~L., 1986, ApJ, 300, 216 


\bibitem[\protect\citeauthoryear{Marscher}{2014}]{2014ApJ...780...87M} 
Marscher A.~P., 2014, ApJ, 780, 87 

\bibitem[\protect\citeauthoryear{Mastichiadis 
\& Kirk}{2002}]{2002PASA...19..138M} Mastichiadis A., Kirk J.~G., 2002, PASA, 19, 138

\bibitem[\protect\citeauthoryear{Miller 
\& Green}{1983}]{1983BAAS...15..957M} Miller H.~R., Green R.~F., 1983, BAAS, 15, 957

\bibitem[\protect\citeauthoryear{Miller et al.}{1988}]{1988ESASP.281b.303M} 
Miller H.~R., Carini M.~T., Gaston B.~J., Hutter D.~J., 1988, ESASP, 281, 
303 

\bibitem[\protect\citeauthoryear{Miller, Carini, 
\& Goodrich}{1989}]{1989Natur.337..627M} Miller H.~R., Carini M.~T., Goodrich B.~D., 1989, Natur, 337, 627 

\bibitem[\protect\citeauthoryear{Mukherjee et 
al.}{1997}]{1997ApJ...490..116M} Mukherjee R., et al., 1997, ApJ, 490, 116

\bibitem[\protect\citeauthoryear{Nalewajko et 
al.}{2011}]{2011MNRAS.413..333N} Nalewajko K., Giannios D., Begelman M.~C., 
Uzdensky D.~A., Sikora M., 2011, MNRAS, 413, 333

\bibitem[\protect\citeauthoryear{Nilsson et 
al.}{2007}]{2007A&A...475..199N} Nilsson K., Pasanen M., Takalo L.~O., Lindfors E., Berdyugin A., Ciprini S., Pforr J., 2007, A\&A, 475, 199

\bibitem[\protect\citeauthoryear{Osterman et 
al.}{2006}]{2006AJ....132..873O} Osterman M.~A., et al., 2006, AJ, 132, 873 

\bibitem[\protect\citeauthoryear{Padovani 
\& Giommi}{1995}]{1995MNRAS.277.1477P} Padovani P., Giommi P., 1995, MNRAS, 277, 1477

\bibitem[\protect\citeauthoryear{Padovani 
\& Giommi}{1996}]{1996MNRAS.279..526P} Padovani P., Giommi P., 1996, MNRAS, 279, 526 

\bibitem[\protect\citeauthoryear{Pian 
\& Treves}{1993}]{1993ApJ...416..130P} Pian E., Treves A., 1993, ApJ, 416, 130 

\bibitem[\protect\citeauthoryear{Raiteri et 
al.}{2003}]{2003A&A...402..151R} Raiteri C.~M., et al., 2003, A\&A, 402, 151 


\bibitem[\protect\citeauthoryear{Rani et al.}{2011}]{2011MNRAS.417.1881R} 
Rani B., et al., 2011, MNRAS, 417, 1881 

\bibitem[\protect\citeauthoryear{Rector 
\& Perlman}{2003}]{2003AJ....126...47R} Rector T.~A., Perlman E.~S., 2003, AJ, 126, 47

\bibitem[\protect\citeauthoryear{Rector, Gabuzda, 
\& Stocke}{2003}]{2003AJ....125.1060R} Rector T.~A., Gabuzda D.~C., Stocke J.~T., 2003, AJ, 125, 1060

\bibitem[\protect\citeauthoryear{Sagar et al.}{2011}]{2011CSci..101.1020S} 
Sagar R., et al., 2011, CSci, 101, 1020 

\bibitem[\protect\citeauthoryear{Sambruna, Maraschi, 
\& Urry}{1996}]{1996ApJ...463..444S} Sambruna R.~M., Maraschi L., Urry C.~M., 1996, ApJ, 463, 444 

\bibitem[\protect\citeauthoryear{Sambruna et 
al.}{2000}]{2000ApJ...538..127S} Sambruna R.~M., et al., 2000, ApJ, 538, 
127

\bibitem[\protect\citeauthoryear{Sasada et al.}{2010}]{2010PASJ...62..645S} 
Sasada M., et al., 2010, PASJ, 62, 645 

\bibitem[\protect\citeauthoryear{Schlegel, Finkbeiner, 
\& Davis}{1998}]{1998ApJ...500..525S} Schlegel D.~J., Finkbeiner D.~P., Davis M., 1998, ApJ, 500, 525

\bibitem[\protect\citeauthoryear{Stetson}{1987}]{1987PASP...99..191S} 
Stetson P.~B., 1987, PASP, 99, 191

\bibitem[\protect\citeauthoryear{Stetson}{1992}]{1992ASPC...25..297S} 
Stetson P.~B., 1992, ASPC, 25, 297 

\bibitem[\protect\citeauthoryear{Sun et al.}{2014}]{2014ApJ...792...54S} 
Sun Y.-H., Wang J.-X., Chen X.-Y., Zheng Z.-Y., 2014, ApJ, 792, 54 

\bibitem[\protect\citeauthoryear{Thompson, Djorgovski, 
\& de Carvalho}{1990}]{1990PASP..102.1235T} Thompson D.~J., Djorgovski S., de Carvalho R., 1990, PASP, 102, 1235 

\bibitem[\protect\citeauthoryear{Urry 
\& Padovani}{1995}]{1995PASP..107..803U} Urry C.~M., Padovani P., 1995, PASP, 107, 803

\bibitem[\protect\citeauthoryear{Villata 
\& Raiteri}{1999}]{1999A&A...347...30V} Villata M., Raiteri C.~M., 1999, A\&A, 347, 30

 \bibitem[\protect\citeauthoryear{Villata et 
al.}{2004}]{2004A&A...421..103V} Villata M., et al., 2004, A\&A, 421, 103

  \bibitem[\protect\citeauthoryear{Wagner 
\& Witzel}{1995}]{1995ARA&A..33..163W} Wagner S.~J., Witzel A., 1995, ARA\&A, 33, 163 
 

\bibitem[\protect\citeauthoryear{Weeks}{2003}]{2003MPLA...18.2099W} Weeks 
J.~R., 2003, MPLA, 18, 2099 

\bibitem[\protect\citeauthoryear{Xie et al.}{2001}]{2001ApJ...548..200X} 
Xie G.~Z., Li K.~H., Bai J.~M., Dai B.~Z., Liu W.~W., Zhang X., Xing S.~Y., 
2001, ApJ, 548, 200 


\bibitem[\protect\citeauthoryear{Xie et al.}{2002}]{2002MNRAS.329..689X} 
Xie G.~Z., Zhou S.~B., Dai B.~Z., Liang E.~W., Li K.~H., Bai J.~M., Xing 
S.~Y., Liu W.~W., 2002, MNRAS, 329, 689


\bibitem[\protect\citeauthoryear{Xie et al.}{2004}]{2004MNRAS.348..831X} 
Xie G.~Z., Zhou S.~B., Li K.~H., Dai H., Chen L.~E., Ma L., 2004, MNRAS, 
348, 831

\bibitem[\protect\citeauthoryear{Xue 
\& Cui}{2005}]{2005ApJ...622..160X} Xue Y., Cui W., 2005, ApJ, 622, 160 

\end{thebibliography}
\end{document}